\documentclass[epj]{svjour}
\usepackage{graphicx}
\usepackage[all]{xy}
\usepackage{amsmath}
\usepackage{amssymb}
\newcommand{\be}{\begin{equation}}
\newcommand{\ee}{\end{equation}}
\newcommand{\ben}{\begin{eqnarray}}
\newcommand{\een}{\end{eqnarray}}
\newcommand{\bes}{\begin{subequations}}
\newcommand{\ees}{\end{subequations}}

\begin{document}
\title{On supersymmetric Dirac delta interactions}

\author{J. Mateos Guilarte\inst{1}\thanks{guilarte@usal.es}\and {J. M. Mu$\tilde{\rm n}$oz Casta$\tilde{\rm n}$eda\inst{2}\thanks{jose.munoz-castaneda@uni-leipzig.de}\and A. Moreno Mosquera\inst{3}\thanks{ asmorenomosquera@gmail.com}}%
}                     
%
%
\institute{Departamento de F\'{\i}sica Fundamental and IUFFyM, Universidad de Salamanca, Spain \and Institut f\"{u}r Theoretische Physik, Universit\"{a}t Leipzig, Germany. \and Facultad Tecnol\'{o}gica. Universidad Distrital Francisco Jos\'{e} de Caldas, Colombia.}

\date{Received: date / Revised version: date}
%
\abstract{
In this paper we construct $\mathcal{N}=2$ supersymmetric (SUSY) quantum mechanics over several configurations of Dirac-$\delta$ potentials from one single delta to a Dirac \lq\lq comb \rq\rq. We show in detail how the building of supersymmetry on potentials with delta interactions placed in two or more points on the real line requires the inclusion of quasi-square wells. Therefore, the basic ingredient of a supersymmetric Hamiltonian containing two or more Dirac-$\delta$s is the singular potential formed by a Dirac-$\delta$ plus a step ($\theta$) at the same point. In this $\delta/\theta$ SUSY Hamiltonian there is only one singlet ground state of zero energy annihilated by the two supercharges or a doublet of ground states paired by supersymmetry of positive energy depending on the relation 
between the Dirac well strength and the height of the step potential. We find a scenario of either unbroken supersymmetry with Witten index one or supersymmetry breaking when there is one \lq\lq bosonic\rq\rq and one \lq\lq fermionic\rq\rq ground state such that the Witten index is zero. We explain next
the different structure of the scattering waves produced by three $\delta/\theta$ potentials with respect to the eigenfunctions arising in the non-SUSY case. In particular, many more bound states paired by supersymmetry exist within the supersymmetric framework compared with the non-SUSY problem. An infinite array of equally spaced $\delta$-interactions of the same strength but alternatively attractive and repulsive are susceptible of being promoted to a ${\cal N}=2$ supersymmetric system. The Bloch's theorem for wave functions in periodic potentials prompts a band spectrum also paired by supersymmetry. Self-isospectrality between the two partner Hamiltonians is thus found. Zero energy ground states are the non-propagating band lower edges, which exist in the spectra of both the two diagonal operators forming the SUSY Hamiltonian. We find that for the SUSY Dirac \lq\lq comb\rq\rq the naif Witten index is zero but supersymmetry is unbroken.}
%
\maketitle


\maketitle

\section{Introduction}\label{intro}
Point or contact interactions modelled by \lq\lq Dirac-$\delta$-potentials\rq\rq give rise to self-adjoint extensions
of free quantum Hamiltonians prompting solvable spectral problems \cite{Albeverio,Bordag-jpa25}. Quantum systems with contact interactions between particles are particularly important in one-dimensional physics giving rise to many-body integrable problems with a very interesting mathematical structure \cite{Lieb,Yang}. In this paper we focus on the analysis of the quantum dynamics of one-body moving on the real line under the influence of several arrays of Dirac potential attractive or repulsive interactions, the one-particle problem behind the dynamics of many-body Yang-Lieb-Liniger systems. Dirac-$\delta$ potentials are interesting also in solid state physics. A Dirac comb formed for infinite $\delta$-walls of identical strength equally spaced, for instance, is the idealization \cite{Cervero} of the Kronig-Penney model \cite{Kronig,Brown} describing electric conductivity in a periodic crystal. 

Recently self-adjoint extensions of free quantum Hamiltonians defined over bounded domains have been studied in the context of the vacuum properties of quantum field theories defined over bounded domains \cite{asoreymunoz-npb874,asoreymunoz-jpa41,asoreymunoz-jpa40,Munoz-Castaneda-prd2013}. In this spirit the contact interactions represent quantum boundary conditions in the formalism developed in reference \cite{aim} and generalised to quantum field theory in references \cite{asoreymunoz-npb874,asoreymunoz-jpa41,asoreymunoz-jpa40}.

In particular, we shall study scattering processes of one-particle systems with one degree of freedom moving in several configurations of Dirac $\delta$-potentials in the framework of supersymmetric quantum mechanics. The idea is the posterior use of the results collected in the calculation of quantum vacuum interactions induced by two kinds of fluctuations between two or more compact $\delta$ objects. Wave functions in one-dimensional supersymmetric quantum mechanics are Pauli two-component spinors, offering room to describe two kinds of fluctuations. There is more structure in this kind of system, particularly the existence of supercharges. Supersymmetric Quantum Mechanics
was invented by Witten in the seminal reference \cite{Witten} with the aim of distinguishing in a framework as simple as possible between systems having singlet ground states of zero energy annihilated by the supercharges and those exhibiting doublet ground states of positive energy paired by the supercharges.

The structure of a ${\cal N}=2$ supersymmetric quantum mechanical system is constructed from a pair of \lq\lq bosonic\rq\rq
canonically conjugated variables, typically the position and momentum operators, supplemented  by one \lq\lq creation\rq\rq and one \lq\lq annihilation\rq\rq fermionic operators space-time independents, see e.g. \cite{Wipf,Wipf1}. 
 In particular, the literature on this problem starting from a Dirac $\delta$-interaction abound. We mention, for instance, the papers \cite{Lboya,Boya1,Boya2,Diaz}, although SUSY Dirac potentials have been selected as pedagogical examples in supersymmetric quantum mechanics in many other references. Recently, a curious mixture of polynomial and Dirac deltas has been extended to supersymmetric quantum mechanics in \cite{Silva}, whereas the intriguing concept of hidden bosonic supersymmetry has been shown to arise also for Dirac $\delta$-interactions, \cite{Plyushchay,Jakubsky}. Our goal in this paper is the construction of ${\cal N}=2$ supersymmetric quantum mechanics for configurations of $N$ $\delta$-interactions on a line. The number of bound states in a configuration of $N$ $\delta$ wells is potentially greater than the unique ground state existing in one $\delta$-well depending the strengths and the separation between $\delta$s (see reference \cite{Munoz-Castaneda-prd2013}). Therefore, it will be convenient to analyse first the bound state dependence on the couplings and the well interdistances in a configuration of $N$ $\delta$s. The reason is that in SUSY Quantum Mechanics the bound states arise in degenerate in energy pairs linked through the supercharges. Remarkably, we shall show that the bound state pairing requires the appearance of quasi-square wells between the attractive $\delta$-interactions such that the number of ground states augments with respect to the non-SUSY case. The general pattern when the $N$ $\delta$-interactions are attractive is technically more involved but qualitatively clear: there is always a zero mode (either bosonic or fermionic) and there are always pairs of bosonic/fermionic degenerate bound states, the number of which increases with $N$.

A very specific arrangement of $\delta$s is simpler to deal with but rich enough to provide interesting information. An array of $N$ equally spaced $\delta$s of the same strength but alternatively attractive and repulsive admit a supersymmetric formulation without additional wells giving rise to an spectral problem as close as possible to twice the spectrum found in identical non-SUSY array. Only the threshold of the continuous spectrum is displaced to push the ground state energy to zero and in one of the two partner Hamiltonians wells are replaced by walls and viceversa. In this paper we shall study this array in the $N\to +\infty$ limit. We shall thus shall address a Dirac comb with alternating white and black teeth, which is accordingly a periodic potential causing a band spectrum that is easily computable. By comparison, the supersymmetric extension of the ordinary Dirac comb, teeth of a single color, is much more complicated. Knowledge of the ground state allows the identification of a superpotential which produces a partner Hamiltonian piecewise formed by P$\ddot{\rm o}$sch-Teller potentials with $\delta$-interactions at the junctions, see \cite{Oeftiger}. The eigenfunctions of the the partner operators clearly differ. The supersymmetric Lam\'e system, however, is realized through two partner isospectral Hamiltonians, a situation described in the references \cite{Sukhatme,Khare2,Dunne,Mannix,Nieto}. Precisely this kind of SUSY spectrum arises in the special array of infinite alternating $\delta$-interactions mentioned above. It is curious that the same spectrum emerges in a second-order SUSY extension of the Lam\'e equation, see \cite{Nieto}, and even  a two-dimensional version of the same construction, see \cite{Ioffe}. 
 
The organization of this paper is as follows: Section \ref{s2} is devoted to the general formulation of the structure of a dynamical system in ${\cal N}=2$ Supersymmetric Quantum Mechanics. In particular we compile some useful formulas that are not common amog the reviews in Supersymmetric Quantum mechanics. In Section \ref{s3} the spectrum of the ensuing supersymmetric Hamiltonian is described for a general superpotential as well as the relationships between the scattering amplitudes of the continuous spectrum of the superpartner Hamiltonians. It is also explained how the positive energy bound states are paired through the supercharges but the zero energy ground states are singlets. A reguralization of the Witten index due to the continuous spectra is offered. In Section \ref{s4} we study the supersymmetric generalisation of one $\delta$-potential, given by a potential in which the Dirac-$\delta$ appears together with a step potential at the same point. This potential is the basic building block of other models to come in the following Sections. It is explained in which circumstances this simple system exhibits spontaneous supersymmetry breaking. Supersymmetric quantum systems of finite arrays of $\delta$s and steps are addressed in Sections \ref{s5}, and \ref{s6}. Section \ref{s8} extends the former calculations to an array of infinite $\delta$-potentials such that the subtleties of the interplay between band spectra and supersymmetry are discussed. Finally, we offer a summary and outlook in Section \ref{s9}.

\section{$\mathcal{N}=2$ extended supersymmetric quantum mechanics: Systems with one degree of freedom}\label{s2}

\subsection{$\mathcal{N}=2$ supersymmetric quantum mechanical systems with one \lq\lq bosonic\rq\rq degree of freedom}
Following the standard construction as 
described e.g. in the references \cite{Khare,coop-ap1983,cromb-ap1983}, we build $\mathcal{N}=2$ supersymmetry on a quantum mechanical Hamiltonian system of one degree of freedom e. g. $\hat{x}$, the position operator of a quantum particle moving on a line, as a \lq\lq bosonic\rq\rq operator in a QFT in $0$-spatial dimensions {\footnote{Of course this terminology has nothing to do with the statistics of an ensemble of several particles. Rather, only the commutation properties of the operators are referred to. Similar considerations apply to the fermionic degree of freedom to come.}}. The promotion to the ${\cal N}=2$ supersymmetric category of this set up is achieved by adding one \lq\lq fermionic\rq\rq degree of freedom characterized by the pair of nilpotent operators $\hat{\psi}$ and $\hat{\psi}^{\dagger}$:
$(\hat{\psi})^2=(\hat{\psi}^\dagger)^2=0$ having physical dimensions of $[\hat{\psi}_k]=M^{-\frac{1}{2}}$ \footnote{$[\mathcal{O}]$ denotes the physical dimensions of the ${\mathcal O}$ observable.}.
Nilpotency demands the anticommutation rules
\begin{equation}
\{\hat{\psi},\hat{\psi}\}=0=\{\hat{\psi}^{\dagger},\hat{\psi}^{\dagger}\},\ \ \ \{\hat{\psi},\hat{\psi}^{\dagger}\}=\frac{1}{m}, \label{rac}
\end{equation}
between the $\hat{\psi}$ and  $\hat{\psi}^{\dagger}$ operators as the canonical quantization rules, together with $[\hat{x},\hat{\psi}]=[\hat{x},\hat{\psi}^\dagger]=0$, involving the fermionic variables.

From these ingredients the supercharges and the supersymmetric Hamiltonian are defined as follows:
\begin{equation}
\hat{Q}=i\hat{\psi}\left(\hbar\frac{d}{d x}+\frac{d W}{d x}\right)=i\hat{\psi}\hat{D}, \ \ \ \hat{Q}^\dagger=i\hat{\psi}^\dagger\left(\hbar\frac{d}{d x}-\frac{d W}{d x}\right)=i\hat{\psi}^\dagger\hat{D}^\dagger, \label{sc1D}
\end{equation}
\begin{equation*}
\hat{H}=-\frac{1}{2m}\left(\hbar\frac{d}{d x}+\frac{d W}{d x}\right)\left(\hbar\frac{d}{d x}-\frac{d W}{d x}\right)-\hbar\hat{\psi}^\dagger\hat{\psi}\frac{d^2W}{d x^2}, \label{oh1d}
\end{equation*}
Here{\footnote{Dealing with supersymetric quantum systems with delta interactions we shall use de facto continuous functions having no continuous first derivatives as superpotentials. Their first derivatives will have finite discontinuities at a discrete set of points in such a way that the second derivatives will be distributions of Dirac delta type. }} $W(x): \mathbb{R} \, \rightarrow \, \mathbb{R}$, is the \lq\lq superpotential\rq\rq. The reason for the name is obvious: the potential energies, responsible for the different interactions
in $\hat{H}$ are determined from $W(x)$. Recalling the representation $\hat{x}=x$, $\hat{p}=-i\hbar\frac{d}{dx}$ of the bosonic variables one can easily check from these definitions and the quantization rules (\ref{rac}) that $\hat{Q}$, $\hat{Q}^\dagger$, and $\hat{H}$ close the superalgebra:
\begin{equation}
\{\hat{Q},\hat{Q}^\dagger\}=2\hat{H} \quad , \quad \{\hat{Q},\hat{Q}\}=\{\hat{Q}^\dagger,\hat{Q}^\dagger\}=0 \, \, \, \label{aqqd} \, .
\end{equation}

\subsection{Clifford algebra representation: supersymmetric states as Pauli spinors}
The fermionic anticommution relations \eqref{rac} are isomorphic to the Clifford algebra relations in $\mathbb{R}^2$. Thus, the Fermi
operators can be represented by means of Pauli matrices. Opposite to the Heisenberg algebra the Clifford algebra admits irreducible finite dimensional representations. The simplest one applicable in our case consists on identifying the vacuum and the one fermion  state with the basic Pauli spinors:
\begin{equation*}
\left|0\right\rangle=\left( \begin{array}
{c}%
1 \\
0
\end{array} \right) \ \ \  , \ \ \ \left|1\right\rangle=\left( \begin{array}
{c}%
0 \\
1
\end{array} \right). \label{repestbase} \, \, ,
\end{equation*}
whereas $\hat{\psi}\left|0\right\rangle=0=\hat{\psi}^\dagger\left|1\right\rangle$ and the equations \eqref{rac} are realized by the $2\times 2$ matrices
$$
\hat{\psi}=\frac{1}{\sqrt{m}}\left( \begin{array}
[c]{cc}%
0 & 1\\
0 & 0
\end{array} \right),\ \ \ \hat{\psi}^{\dagger}=\frac{1}{\sqrt{m}}\left( \begin{array}
[c]{cc}%
0 & 0\\
1 & 0
\end{array} \right).
$$
The Fermi/Bose number operators and their Klein counterparts are respectively:
$$
\hat{F}=\left( \begin{array}
[c]{cc}%
0 & 0\\
0 & 1
\end{array} \right), \\\ \hat{B}=\left( \begin{array}
[c]{cc}%
1 & 0\\
0 & 0
\end{array} \right), \ \ \ \hat{K}_F=\left( \begin{array}
[c]{cc}%
1 & 0\\
0 & -1
\end{array} \right), \\\ \ \ \ \hat{K}_B=\left( \begin{array}
[c]{cc}%
-1 & 0\\
0 & 1
\end{array} \right).
$$
\\
The supercharges in (\ref{sc1D}) become $2\times 2$ matrix first-order differential operators and the Hamiltonian is a $2\times 2$ diagonal matrix of Schr$\ddot{\rm o}$dinger operators:
\begin{equation*}
\hat{Q}=\frac{i}{\sqrt{m}}\left( \begin{array}
[c]{cc}%
0 & \hat{D}\\
0 & 0
\end{array} \right), \ \ \ \hat{Q}^{\dagger}=\frac{i}{\sqrt{m}}\left( \begin{array}
[c]{cc}%
0 & 0\\
\hat{D}^{\dagger} & 0
\end{array} \right), \qquad 
\hat{H}=\left( \begin{array}
[c]{cc}%
\hat{H}_0& 0\\
0 & \hat{H}_1
\end{array} \right) 
\end{equation*}
\begin{equation*}
\hat{H}_0=-\frac{\hat{D}\hat{D}^\dagger}{2m}=-\frac{\hbar^2}{2m}\frac{d^2}{dx^2}+\frac{1}{2m}{W^{\prime}}^2+\frac{\hbar}{2m}W^{\prime\prime}, 
\end{equation*}
\begin{equation*}
\hat{H}_1=-\frac{\hat{D}^\dagger\hat{D}}{2m}=-\frac{\hbar^2}{2m}\frac{d^2}{dx^2}+\frac{1}{2m}{W^{\prime}}^2-\frac{\hbar}{2m}W^{\prime\prime}.
\end{equation*}
Finally, the supersymmetric states become simply Pauli spinor wave functions: $
\Psi(x)=\langle x |\Psi\rangle=\left(\begin{array}{c} \psi_0(x) \\ \psi_1(x) \end{array}\right)$.

\section{The spectrum of the supersymmetric Hamiltonian}\label{s3}
The spectrum of a supersymmetric Hamiltonian $\hat{H}|\Psi_E\rangle=E|\Psi_E\rangle$ is non-negative:
\begin{equation*}
E=\frac{\langle\Psi_E|\hat{H}|\Psi_E\rangle}{\langle\Psi_E|\Psi_E\rangle}=\frac{1}{2\langle\Psi_E|\Psi_E\rangle}\left(\langle\hat{Q}\Psi_E|\hat{Q}\Psi_E\rangle +\langle\hat{Q}^\dagger\Psi_E|\hat{Q}^\dagger\Psi_E\rangle\right)\geq 0 \quad .
\end{equation*}
The energy eigenstates are also eigenstates of the Fermi number operator of two types
\begin{eqnarray*}
\hat{F}|\Psi_E\rangle_0&=&\hat{F}\left(f_{0E}|0\rangle\right)=0|\Psi_E\rangle_0 \quad , \quad \hat{K}_F|\Psi_E\rangle_0=|\Psi_E\rangle_0 \\
\hat{F}|\Psi_E\rangle_1&=&\hat{F}\left(f_{1E}|1\rangle\right)=1|\Psi_E\rangle_1 \quad , \quad \hat{K}_F|\Psi_E\rangle_1=-|\Psi_E\rangle_1
\end{eqnarray*}
that we shall call \lq\lq bosonic\rq\rq and \lq\lq fermionic\rq\rq .

\subsection{Bound states: singlets and doublets}

There are two types of bound states in a supersymmetric quantum mechanical system. These can be classified according to their energies (see ref. \cite{Khare} for more details):
\begin{enumerate}
\item \underline{Zero modes}. These states belong to the kernel of the supercharges. It is clear that $\hat{Q}$ also annihilates the bosonic zero modes and $\hat{Q}^\dagger$ does the same with the fermionic zero modes. With only one degree of freedom there are at most
two states of this type, each of them forming one \underline{singlet} state (short multiplet) of the supersymmetry algebra.

\item \underline{Pairs of positive energy bound states}.  The energy potentials
\begin{equation*}
V_s(x)=\frac{1}{2m}{W^{\prime}}^2+(-1)^s\frac{\hbar}{2m}W^{\prime\prime} \quad , \quad \lim_{x\to\pm\infty}V_s(x)=v_\pm^2(s) \, \, , \, \, s=0,1
\end{equation*}
may give rise to $j=1,2, \cdots , n_b$  positive energy bound eigenstates in the supersymmetric Hamiltonian that come in pairs and form doublets (long multiplets) of the
supersymmetry algebra (see \cite{Khare}). If $v_\pm^2=\infty$ the natural number $n_b$ goes to $\infty$, i.e., the spectrum of $\hat{H}$ is purely discrete. Even though in the most general case $v_\pm^2(s)$ depends on $s$ we will assume from now onwards that it does not depend on $s$.
\end{enumerate}

\subsection{Supersymmetric scattering amplitudes}
If $v_\pm^2<+\infty$ there is a threshold for scattering eigenstates at $E={\rm min}(v_-,v_+)$. Let us consider a plane wave basis in the completion of $L^2(\mathbb{R})$:
$\hat{p}|k\rangle=\hbar k|k\rangle$, $f_k(x)=\langle x|k\rangle=\frac{1}{\sqrt{2\pi}}e^{ikx}$. Defining $\hbar k_\pm=\sqrt{E-v_\pm^2}$ and
considering energies $E\geq {\rm max}(v_-,v_+)$ the scattering eigenstates of the Hamiltonian have the asymptotic form:
\begin{eqnarray*}
&&|\Psi_E^r\rangle_0= \left\{\begin{array}{ll} |k_-\rangle_0 +\rho_0^r(E) |-k_-\rangle \\ \sigma_0^r(E)|k_+\rangle \end{array}\right. \hspace{0.2cm} , \hspace{0.2cm}
|\Psi_E^r\rangle_1= \left\{\begin{array}{ll} |k_-\rangle_1 +\rho_1^r(E) |-k_-\rangle \\ \sigma_1^r(E)|k_+\rangle \end{array}\right.  \begin{array}{c} , \ \ x<<0 \\ , \ \ x>>0 \end{array}\\
&&|\Psi_E^l\rangle_0= \left\{\begin{array}{ll} \sigma_0^l(E)|-k_-\rangle \\ |-k_+\rangle_0 +\rho_0^l(E) |k_+\rangle  \end{array}\right. \hspace{0.2cm} , \hspace{0.2cm}
|\Psi_E^l\rangle_1= \left\{\begin{array}{ll} \sigma_1^l(E)|-k_-\rangle \\ |-k_+\rangle_1 +\rho_1^l(E) |k_+\rangle  \end{array}\right.   \begin{array}{ll} , \ \ x<<0 \\ , \ \ x>>0 \end{array}
\end{eqnarray*}
corresponding to the right moving waves in the first row and the left moving ones in the second row, both bosonic and fermionic.
Except when the threshold lies exactly at $0$, $v_\pm=0$, these states also form doublets of the supersymmetry algebra:
\begin{equation}
|\Psi_E^r\rangle_1=\hat{Q}^\dagger|\Psi_E^r\rangle_0 \quad , \quad |\Psi_E^l\rangle_1=\hat{Q}^\dagger|\Psi_E^l\rangle_0 \quad . \label{scatp}
\end{equation}
Recalling that the supercharge has the form $\hat{Q}^\dagger=-\hat{\psi}^\dagger\left(\hat{p}+i W^\prime(x)\right)$ and taking into account (\ref{scatp}) the following identities between
the scattering amplitudes hold \cite{Khare,Wipf,cooper-pla1988}:
\begin{eqnarray}
\sigma_1^r(E)&=&\frac{i\hbar k_+-v_+}{i\hbar k_-+v_-}\sigma_0^r(E) \quad , \quad \rho_1^r(E)=-\frac{i\hbar k_--v_-}{i\hbar k_-+v_-}\rho_0^r(E) \label{scatar}\\
\sigma_1^l(E)&=&\frac{i\hbar k_-+v_-}{i\hbar k_+-v_+}\sigma_0^l(E) \quad , \quad \rho_1^l(E)=\frac{i\hbar k_++v_+}{i\hbar k_+-v_+}\rho_0^r(E) \label{scatal} \quad .
\end{eqnarray}
In the derivation of the formulas (\ref{scatar}) and (\ref{scatal}) we have assumed very mild conditions for $W(x)$:
\begin{equation*}
\lim_{x\to +\infty}W^\prime (x)=v_+ \quad , \quad \lim_{x\to -\infty}W^\prime (x)=-v_- \quad , \quad \lim_{x\to\pm\infty}W^{\prime\prime}(x)=0 \quad .
\end{equation*}
Therefore, supersymmetry shows itself in the fact that the moduli of the scattering amplitudes are invariant:
\begin{equation*}
\left|\sigma_0^r(E)\right|=\left|\sigma_1^r(E)\right|;  \, \left|\rho_0^r(E)\right|=\left|\rho_1^r(E)\right| ; \,  \left|\sigma_0^l(E)\right|=\left|\sigma_1^l(E)\right|; \,  \left|\rho_0^l(E)\right|=\left|\rho_1^l(E)\right| \,  .
\end{equation*}
Moreover, as in the non-supersymmetric systems there is conservation of probability density fluxes:
\begin{eqnarray*}
&&\frac{k_+}{k_-}\left|\sigma_0^r(E)\right|^2+\left|\rho_0^r(E)\right|^2=\frac{k_+}{k_-}\left|\sigma_1^r(E)\right|^2+\left|\rho_1^r(E)\right|^2=1 \\
&&\frac{k_-}{k_+}\left|\sigma_0^l(E)\right|^2+\left|\rho_0^l(E)\right|^2=\frac{k_-}{k_+}\left|\sigma_1^l(E)\right|^2+\left|\rho_1^l(E)\right|^2=1 \quad .
\end{eqnarray*}

\subsection{Spontaneous supersymmetry breaking}

If the spectrum of $\hat{H}$ is purely discrete the Witten index $I_W$ is defined as the difference between the number of bosonic zero modes $z_0=\dim {\rm ker}(\hat{H}_0)$ and the number of fermionic zero modes $z_1=\dim {\rm ker}(\hat{H}_1)$:  $I_W=z_0-z_1$ (see references \cite{Witten1,Freund}). When there is spontaneous supersymmetry breaking $I_W=0$\footnote{The reverse is not true: $I_W=0$ does not imply that there is spontaneous supersymmetry breaking if $z_0,z_1\neq 0$.}. In the case of mixed
spectrum -discrete plus continuous- it is convenient to introduce the following regularization of the Witten index:
\begin{equation*}
I_W=\lim_{t\to 0}{\rm Tr}_{L^2}\, \hat{K}_F \, e^{-\frac{t\hat{H}}{\hbar}} \label{witin}
\end{equation*}
 where $\hat{K}_F$ is the the fermionic Klein operator and $t$ can be understood as Euclidean time. In particular, if $v_+=v_-=v$ such that $k_+=k_-=k$ and $\sigma^r(k)=\sigma^l(k)$, setting $\hbar=1$, we obtain:
\begin{equation*}
I_W=z_0-z_1+\lim_{t\to 0}\int_{-\infty}^\infty \, dk \, \left(\varrho_0(k)-\varrho_1(k)\right){\rm exp}\left[-t(k^2+v^2)\right] \quad .
\end{equation*}
The spectral densities are defined from the phase shifts when the system is restricted to a finite interval of very long length $L$ and subjected to periodic boundary conditions in the form:
\begin{equation*}
\varrho_s(k)=\frac{L}{2\pi}+\frac{1}{2\pi}\frac{d\delta_s}{dk}(k) \, \, , \, \, \, \delta_s(k)=\delta_{s+}(k)+\delta_{s-}(k) \, \, , \, \, s=0,1
\end{equation*}
The supersymmetric phase shifts are:
\begin{equation*}
\delta_{s\pm}(k)=\frac{1}{2i}\log\left[\sigma_s(k)\pm\sqrt{\rho_s^r(k)\rho_s^l(k)}\right];\,\,s=0,1 
\end{equation*}
whereas the supersymmetric scattering amplitudes are related as follows:
\begin{equation*}
\sigma_1(k)=\frac{ik-v}{ik+v}\sigma_0(k) \, \, , \, \, \rho_1^{r,l}(k)=-\frac{ik-v}{ik+v}\rho^{r,l}_0(k) .
\end{equation*}
Therefore,
\begin{eqnarray*}
&& \varrho_0(k)-\varrho_1(k)=-\frac{1}{2\pi i}\frac{d}{dk}\left(\log\left[\frac{ik-v}{ik+v}\right]\right)=\frac{v}{\pi}\cdot\frac{1}{k^2+v^2}\\
&&\lim_{t\to 0}\left(\frac{v}{\pi}\int_{-\infty}^\infty\, {\rm exp}\left[-t(k^2+v^2)\right]\frac{dk}{k^2+v^2}\right)=\lim_{t\to 0}\left(1-{\rm Erf}[v\sqrt{t}]\right)=1
\end{eqnarray*}
and finally we obtain: $I_W=z_0-z_1+1$. Within this rather standard heat trace regularization in scattering problems the Witten index is shifted by one in virtue of the difference between the spectral densities of the continuous spectra of the SUSY partner Hamiltoninans $\widehat H_0$ and $\widehat H_1$. This result looks like a simple application of the index theorem for the Dirac operator on non-compact manifolds describig the chiral anomaly not only in terms of the first Chern class but also picking some contribution from the scattering chiral spinors phase shifts, see Reference \cite{for-npb1987}

\section{The Dirac $\delta$/step potential}\label{s4}

\subsection{Analysis of the non-supersymmetric problem}
In ordinary quantum mechanics the Hamiltonian (in a system of units such that
$\hbar=1, m=\frac{1}{2}$)
for a particle moving on the real line and confronting one step plus a Dirac $\delta$ potential at the same point (the origin) is:
\[
\hat{H}=-\frac{d^2}{dx^2}+\mu \delta(x)+\frac{g}{2}\varepsilon(x)+\frac{g}{2}.
\]
Here $\delta(x)$ and $\varepsilon(x)$ are respectively the Dirac $\delta$ and sign distributions, whereas $\mu$
is a parameter of dimensions $L^{-1}$ setting the strength of the $\delta$ potential and $g$, with dimensions $L^{-2}$,
measures the height of the step. The spectral problem is equivalent to the Schr$\ddot{\rm o}$dinger equation
\begin{equation}
-\psi^{\prime\prime}+\left[\mu\delta(x)+\frac{g}{2}\varepsilon(x)+\frac{g}{2}\right]\psi(x)=E\psi(x), \label{esdp}
\end{equation}
together with the matching conditions at the origin, 
\begin{equation}
\lim_{x\to 0^+}\psi(x)=\lim_{x\to 0^-}\psi(x) \quad , \quad \lim_{x\to 0^+}\psi^\prime(x)-\lim_{x\to 0^-}\psi^\prime(x)=\mu \lim_{x\to 0^-}\psi(x) \label{matcon}
\end{equation}
to be held by the scattering solutions.

One must keep in mind that the SUSY spectrum is non-negative. Therefore to obtain a supersymmetric extension of the Dirac-$\delta$ potential one must add a step term $\frac{g}{2}\varepsilon(x)+\frac{g}{2}$ to the potential to push the negative energy levels from the Dirac-$\delta$ to positive energy values. This is a fundamental requirement of SUSY.
\subsubsection{Scattering waves}
 The continuity/discontinuity conditions (\ref{matcon}) define the self-adjoint extension  of the free particle Hamiltonian equivalent to the Dirac $\delta$ potential and the fact that $k_-=k=\sqrt{E}$ on the negative half-line whereas $k_+=p=\sqrt{E-g}$ on the positive half-line is due to the jump in energy at the step $\varepsilon$-potential. Together
 with the scattering wave ansatz (waves incoming towards the $\delta$/$\varepsilon$-potential either from the left, labeled by the superscript $r$, or from the right, labeled by $l$)
 \begin{equation*}
\psi^r(x,E)=\left\{ \begin{array}{ll}
e^{ikx}+\rho^r (E)e^{-ikx} & , \ \ x\rightarrow -\infty\\
\sigma^r(E) e^{ipx} & , \ \ x\rightarrow \infty\\
\end{array} \right. \quad , \quad
\psi^l(x,E)=\left\{ \begin{array}{ll}
\sigma^l(E) e^{-ikx} & , \ \ x\rightarrow -\infty\\
e^{-ipx}+\rho^l(E)e^{ipx} & , \ \ x\rightarrow \infty\\
\end{array} \right.
\end{equation*}
the matching conditions (\ref{matcon}) allow to identify the scattering amplitudes
\begin{eqnarray*}
\sigma^r(E) &=& \frac{2 k}{k+p+i \mu } \quad , \quad \rho^r(E)=\frac{k-p-i \mu }{k+p+i \mu }\\
  \sigma^ l(E)&=&  \frac{2 p}{k+p+i \mu }\quad , \quad \rho^l(E)=\frac{-k+p-i \mu }{k+p+i \mu }
\end{eqnarray*}
by solving an algebraic linear system of two equations in two unknowns.
The conservation of the total probability density flux is manifest:
\begin{eqnarray*}
&& \frac{p}{k}|\sigma^r(E)|^2+ |\rho^r(E)|^2=\frac{k}{p}|\sigma^l(E)|^2+ |\rho^l(E)|^2=\\
&&  \frac{p}{k}\left|\frac{2 k}{k+p+i \mu }\right|^2+\left|\frac{k-p-i \mu }{k+p+i \mu }\right|^2=\frac{k}{p}\left|\frac{2 p}{k+p+i \mu }\right|^2
+ \left|\frac{-k+p-i \mu }{k+p+i \mu }\right|^2=1
\end{eqnarray*}

\subsubsection{Bound/anti-bound states}
 The poles of the transmission amplitudes $\sigma^r$ and $\sigma^l$, the purely imaginary solutions of the equation $k+p+i \mu=0=i(\kappa + \mathcal{\eta}+\mu)$, correspond to either bound ($\kappa>0$) or anti-bound\footnote{It is of note that even though the anti-bound states are characterised by negative pure imaginary poles of the transmission amplitude they are not physical states. Even more they are not a point of the spectrum} ($\kappa<0$) solutions of the Schr\"odinger equation \eqref{esdp} of the form 
 \begin{equation}
\psi(x)=N \ [e^{\kappa x}\theta(-x)+e^{-\mathcal{\eta} x}\theta(x)] \label{wfdp0}
\end{equation}
where $\theta(x)$ is the step Heaviside function if $k=i\kappa$, $p=i\mathcal{\eta}$ and $\kappa\in\mathbb{R}$, $\mathcal{\eta}\in\mathbb{R}$. The existence of the bound state is a more subtle problem than in the $g=0$ case. Setting, e.g., $g>0$
($g<0$ yields a completely analogous situation exchanging $\kappa$ and $\mathcal{\eta}$), there are several possibilities:
\begin{enumerate}
\item $\mu>0$: The $\delta$-potential is repulsive and the right-to-left and left-to-right transmission amplitudes $\sigma^r$ and $\sigma^l$ have a unique purely imaginary pole in $k+p=i(\kappa + \mathcal{\eta})=-i\mu$. From the relation $\kappa^2-\mathcal{\eta}^2=-g$ the pole in terms of $\kappa$ and $\mathcal{\eta}$ can be separately identified:
    $\kappa=-\frac{1}{2}(\mu-\frac{g}{\mu})$ and $\mathcal{\eta}=-\frac{1}{2}(\mu+\frac{g}{\mu})$. Given that  $\mathcal{\eta}<0$ the associated wave function (\ref{wfdp0}) is non-normalizable. Accordingly, this pole corresponds to an anti-bound state.
\item $\mu<0$, \ $|\mu|^{2}>g$: The $\delta$-potential is attractive  and the pole of $\sigma^r$ and $\sigma^l$ is found at: $\kappa=\frac{1}{2}(|\mu|-\frac{g}{|\mu|})$  and $\mathcal{\eta}=\frac{1}{2}(|\mu|+\frac{g}{|\mu|})$. Thus, both $\kappa>0$ and $\mathcal{\eta}>0$ are positive and a true (normalizable) bound state corresponds to the pole in this range of the parameters.
    \\
    If $\mu<0$ \underline{and} $|\mu|^{2}>g$ there exists a bound state of energy $E=-\kappa^2=-\frac{1}{4}(|\mu|-\frac{g}{|\mu|})^{2}$ and normalized wave function:
    \begin{equation}
\psi(x)=\sqrt{\frac{|\mu|^4-g^2}{2|\mu|^3}} \left(e^{\frac{|\mu|^2-g}{2|\mu|} x}\theta(-x)+e^{[-\frac{|\mu|^2+g}{2|\mu|} x}\theta(x)\right) \label{wfdp}
\end{equation}
\item $\mu<0$, \ $|\mu|^{2}<g$: The difference is that now $\kappa=\frac{|\mu|^2-g}{2|\mu|}<0$ and the wave function (\ref{wfdp}) becomes non-normalizable (the norm is imaginary). The pole in this range of the parameters gives an anti-bound state.
\end{enumerate}
In summary, there is one bound state if and only if the $\delta$ potential is an attractive well, $\mu<0$, strong enough, $|\mu|^2>g$, to
overcome the repulsion at the step $g$.
\subsection{$\mathcal{N}=2$ supersymmetric Dirac $\delta$/$\varepsilon$ Hamiltonian}
Let us choose
\begin{equation}
W(x)=\frac{\mu}{2}\left|x\right|+\frac{g}{2\mu}x, \label{wdps}
\end{equation}
as the superpotential. The partner scalar Hamiltonians acting respectively on the sub-spaces of zero ($s=0$) and one ($s=1$) Fermi number are:
\begin{equation*}
\hat{H}_s\equiv\left.\hat{H}\right|_{\mathcal{SH}_s}=-\frac{d^{2}}{dx^{2}}+\frac{g}{2}\epsilon(x)+\frac{\mu^{2}}{4}+\frac{g^{2}}{4\mu^{2}}+(-1)^s\mu\delta(x)
\end{equation*}
Assuming $\mu>0$ and $g>0$ $V_0$ and $V_1$ correspond to respectively repulsive and attractive Dirac delta potentials of strength $\mu$ plus one step potential of height $g$ shifted over zero energy by the quantity $\frac{1}{4}(\mu-\frac{g}{\mu})^{2}$. Changing the sign of $\mu$ merely exchanges the repulsive/attractive character of the $\delta$ potential between $V_0$ and $V_1$. For fixed $\mu$ changing the sign of $g$ the step is reversed from right to left or vice versa. The key point is that the shift in the energy of the scattering threshold with respect to the non-SUSY threshold exactly pushes the energy of the ground state to zero. This is necessary because the spectrum of a super-symmetric Hamiltonian is non-negative.
\\
The Schr$\ddot{\rm o}$dinger equations to be solved are:
\begin{equation*}
-\frac{d^2{\psi}^{(s)}}{dx^2}+V_s\psi^{(s)}(x)=E^{(s)}\psi^{(s)}(x); \quad s=0,1,
\end{equation*}
together with the continuity of $\psi^{(s)}$ and the discontinuity of $\frac{d{\psi}^{(s)}}{dx}$, at the $x =0$ point, according to the matching conditions (\ref{matcon}).

\subsubsection{Scattering waves}
The scattering solutions of these equations are similar to those of the non-SUSY problem. The only difference is in the relations between momenta and energies, which in this case are:
\begin{equation*}
k^2=E^{(s)}-\frac{1}{4}\left(\mu-\frac{g}{\mu}\right)^{2} \qquad,
 \qquad  p^2=E^{(s)}-\frac{1}{4}\left(\mu+\frac{g}{\mu}\right)^{2} \quad .
\end{equation*}
From these identities it is clear that $E^{(0)}=E^{(1)}=E$ for the scattering solutions as the SUSY algebra demands. The matching conditions (\ref{matcon}), however, on the the left-to-right and right-to-left waves:
\begin{eqnarray*}
\psi^{r(s)}(x,E)&=&\left(e^{ikx}+\rho^r_s(E)e^{-ikx}\right)\theta(-x)+\sigma^r_s(E)e^{ipx}\theta(x)\\
\psi^{l(s)}(x,E)&=&\left(e^{-ipx}+\rho^l_s(E)e^{ipx}\right)\theta(x)+\sigma^l_s(E)e^{-ikx}\theta(-x) \, \, , \, s=0,1
\end{eqnarray*}
give rise to the different scattering amplitudes:
\begin{eqnarray*}
&&\sigma_{s}^r(E)=\frac{2 k}{k+p+i(-1)^s\mu}  , \, \, \rho_{s}^r(E)=\frac{k-p-i(-1)^s\mu}{k+p+ i(-1)^s\mu};\\ 
&& \sigma_{s}^l(E)=\frac{2 p}{k+p+i(-1)^s\mu}  , \, \, \rho_{s}^l(E)=\frac{-k+p-i(-1)^s\mu}{k+p+ i(-1)^s\mu};\, \, s=0,1.
\end{eqnarray*}
\\
These scattering waves are super-symmetric in the sense that, for example, the left-to-right waves in the two sectors are connected in the form: 
\[
\left(\begin{array}{c}0 \\ \psi_{1}^r(x;E)\end{array}\right)\propto \hat{Q}^\dagger \left(\begin{array}{c}\psi_{0}^r(x;E)\\ 0\end{array}\right)
\, \, \, {\rm with} \, \, \,
\hat{Q}^\dagger=\left(\begin{array}{cc}0 & 0 \\ \frac{d}{dx}-\frac{\mu}{2}\varepsilon(x)-\frac{g}{2\mu} & 0 \end{array}\right) \, \, .
\] 
Because,
\begin{eqnarray*}
&&\hspace{-0.8cm}\left( \frac{d}{dx}-\frac{\mu}{2}\varepsilon(x)-\frac{g}{2\mu}\right)\psi_{0}^r(x;E)=\left(ik+\frac{\mu}{2}-\frac{g}{2\mu}\right)\times
\\&&\hspace{-0.8cm}\times
\left\{\left[e^{ikx}+\frac{-ik+\frac{\mu}{2}-\frac{g}{2\mu}}{ik+\frac{\mu}{2}-\frac{g}{2\mu}}\rho_{0}^r(E)e^{-ikx}\right]\theta(-x)
+\frac{i p-\frac{\mu}{2}-\frac{g}{2\mu}}{ik+\frac{\mu}{2}-\frac{g}{2\mu}}\sigma_{0}^r(E)e^{ipx}\theta(x)\right\}
\end{eqnarray*}
$\psi_{1}^r(x;E)$ and $\psi_{0}^r(x;E)$ are paired by supersymmetry if the scattering amplitudes satisfy:
\begin{equation}
\rho_{1}^r=\frac{-i k+\frac{\mu}{2}-\frac{g}{2 \mu}}{i k+\frac{\mu}{2}-\frac{g}{2 \mu}}\rho_{0}^r \quad , \quad
\sigma_{1}^r=\frac{i p-\frac{\mu}{2}-\frac{g}{2 \mu}}{i k+\frac{\mu}{2}-\frac{g}{2\mu}}\sigma_{0}^r\label{igual}
\end{equation}
From the explicit form of the scattering amplitudes above it is not difficult to check that the identities (\ref{igual}) hold{\footnote{Analogous relations for the left-going amplitudes also hold.}}. Even though $\sigma_{0}^r\neq \sigma_{1}^r$ ($\sigma_{0}^l\neq \sigma_{1}^l$) and $\rho_{0}^r\neq \rho_{1}^r$ ($\rho_{0}^l\neq \rho_{1}^l$) the scattering coefficients of the SUSY partner Hamiltonians are identical:
\begin{equation*}
\left|\rho_{0}^r\right|^2=\left|\rho_{1}^r\right|^2=1-\frac{4 k p}{\mu ^2+(k+p)^2}=\left|\rho_{0}^l\right|^2=\left|\rho_{1}^l\right|^2
\end{equation*}
\begin{equation*}
\left|\sigma_{0}^r\right|^2=\left|\sigma_{1}^r\right|^2=\frac{4 k^2}{\mu^2+(k+p)^2} \quad , \quad \left|\sigma_{0}^l\right|^2=\left|\sigma_{1}^l\right|^2=\frac{4 p^2}{\mu^2+(k+p)^2} \quad .
\end{equation*}
We remark that despite $\sigma^r \neq \sigma^l$ there is probability density flux conservation: $\frac{k_\pm}{k_\mp}|\sigma|^2+|\rho|^2= 1$ in the two sectors.
\subsubsection{Bound/anti-bound states}
 Like in the non-SUSY system the pole in the transmission amplitudes of $\hat{H}_0$ $\sigma_{0}^r$ and $\sigma_{0}^l$
 appears in:
 \[
 \kappa_0=-\frac{1}{2}\left(\mu-\frac{g}{\mu}\right) \qquad ,  \quad \mathcal{\pi}_0=-\frac{1}{2}\left(\mu+\frac{g}{\mu}\right) \quad .
 \]
 Because $\kappa_1+\mathcal{\pi}_1=\mu$ is the pole of $\sigma_{1}^r$ and $\sigma_{1}^l$ but still $\kappa_1^2-\mathcal{\pi}_1^2=-g$ the separate values of $\kappa_1$ and $\mathcal{\pi}_1$ at the pole are:
 \[
 \kappa_1=\frac{1}{2}\left(\mu-\frac{g}{\mu}\right) \qquad , \qquad \mathcal{\pi}_1= \frac{1}{2}\left(\mu+\frac{g}{\mu}\right)\quad .
 \]
The bound/anti-bound wave functions are respectively:
\begin{eqnarray*}
\hspace{0.5cm}\psi^{(0)}(x)&=&\sqrt{\frac{g^2-\mu^4}{2\mu^3}} \left({\rm exp}\left[\frac{g-\mu^2}{2\mu} x\right]\theta(-x)+{\rm exp}\left[\frac{\mu^2+g}{2\mu} x\right]\theta(x)\right) \label{wfdps0} \\ \hspace{0.5cm}\psi^{(1)}(x)&=&\sqrt{\frac{\mu^4-g^2}{2\mu^3}} \left({\rm exp}\left[\frac{\mu^2-g}{2\mu} x\right]\theta(-x)+{\rm exp}\left[-\frac{\mu^2+g}{2\mu} x\right]\theta(x)\right)  . \label{wfdps1}
\end{eqnarray*}
There are two possibilities if we set (with no loss of generality) $g>0$:
\begin{enumerate}
\item $\mu^2>g$. There is a unique ground state (\lq\lq bosonic \rq\rq for $\mu<0$ and \lq\lq fermionic \rq\rq for $\mu>0$) due to normalizability conditions of the wave functions $\psi^{(0)}(x)$ and $\psi^{(1)}(x)$. Therefore in this case the Witten index is one  and supersymmetry is not spontaneously broken in these cases.

\item $\mu^2<g$. Neither $\psi^{(0)}(x)$ nor $\psi^{(1)}(x)$ are normalizable. Hence there are no zero modes and supersymmetry is spontaneously broken.
\end{enumerate}

Finally, it is obvious that in the $g\rightarrow 0$ limit we recover the Dirac $\delta$ potential in the supersymmetric version \cite{Boya1,Boya2,Diaz,Plyushchay,Jakubsky,Silva}.

\section{The SUSY double Dirac delta potential}\label{s5}
In this section we analyse the the supersymmetric double Dirac-$\delta$ potential. The analysis to be carried out is the exactly the one carried out for the non-supersymmetric double Dirac-$\delta$ in reference \cite{Munoz-Castaneda-prd2013}. In this reference the hamiltonian 
\begin{equation}
\hat{H}=-\frac{d^2}{dx^2}+V(x)=-\frac{d^2}{dx^2}+\alpha \delta(x+a)+ \beta \delta(x-a) \quad . \label{h2del}
\end{equation}
is analysed in order to study the quantum vacuum interaction energy between two Dirac-$\delta$ potentials. The steps to follow in the study of the double supersymmetric Dirac-$\delta$ potential are:
\begin{itemize}
\item Characterise the scattering states of the system, and the corresponding scattering amplitudes. 
\item Compute the scattering matrix and the analytic function that characterises the poles of the transmission amplitude.
\item Study the poles of the transmission amplitudes to investigate the existence of bound and anti-bound states in terms of the parameters that characterise each of the supersymmetric Dirac-$\delta$s. For the case of two non-supersymmetric Dirac-$\delta$ potentials reference \cite{Munoz-Castaneda-prd2013} shows that depending on the values of the parameters $\alpha a$ and $\beta a$ defined in the Hamiltonian (\ref{h2del}) one can have regions where there are one or two bound states or regions where there are none. Figure 1 in Ref. \cite{Munoz-Castaneda-prd2013} shows how these three regions are distributed in the $\alpha a$-$\beta a$ plane
\end{itemize}

For the case of the supersymmetric double Dirac-$\delta$ we distinguish between different cases depending on the weights of each Dirac-$\delta$ step potential.

In the double Dirac-$\delta$ it is required to add a square well to the non-SUSY potential  in order to ensure the SUSY. In addition the depth of the well is determined by SUSY. As a consequence this square well imposed by SUSY ensures non-negative spectrum which is a fundamental property of SUSY quantum mechanics

\subsection{$\mathcal{N}=2$ supersymmetric double Dirac-$\delta$ Hamiltonian: two $\delta$s of the same strength}

To work the $\mathcal{N}=2$ super-symmetric extension of this system it is convenient to deal separately with the cases of equal and different strength. In the symmetric case we choose the following super-potential:
\begin{equation*}
W(x)=\frac{\alpha}{2}\left|x+a\right|+\frac{\alpha}{2}\left|x-a\right|, \label{wdds}
\end{equation*}
\\
The SUSY partner scalar Hamiltonians are:
\begin{equation}
\hat{H}_s=-\frac{d^2}{dx^2}+(-1)^s\alpha\delta(x+a)+(-1)^s\alpha\delta(x-a)+\frac{\alpha^{2}}{2}\epsilon(x+a)\epsilon(x-a)+\frac{\alpha^{2}}{2},
\end{equation}
being $s=0,1$ the Fermi number. The Schr$\ddot{\rm o}$dinger equations
\begin{equation*}
\hat{H}_0\psi^{(0)}(x)=E^{(0)}\psi^{(0)}(x);\quad \hat{H}_1\psi^{(1)}(x)=E^{(1)}\psi^{(1)}(x)
\end{equation*}
must be solved together with the continuity of $\psi^{(s)}$ and the discontinuity of $\frac{d{\psi}^{(s)}}{dx}$, $s=0,1$, at the points $x =\pm a$. These are the same matching conditions described in references \cite{Munoz-Castaneda-prd2013,Bordag-jpa25}.

\subsubsection{Scattering waves}
Besides of the two $\delta$s at the points $x=\pm a$, with opposite signs of $\alpha$ respectively for $V_0$ and $V_1$, the partner potentials $V_0(x)=\frac{dW}{dx}\frac{dW}{dx}+\frac{d^2W}{dx^2}$ and $V_1(x)=\frac{dW}{dx}\frac{dW}{dx}-\frac{d^2W}{dx^2}$ exhibit an identical square well: $V_0(x)=V_1(x)=\alpha^2 , |x| > a$ (in zones II and III), but $V_0(x)=V_1(x)=0 , |x| < a$ (in zone I). The \lq\lq right-going\rq\rq  and \lq\lq left-going\rq\rq scattering wave functions, both in the bosonic and fermionic sectors, are of the form:
\begin{equation*}
\psi^{r(s)}(x,E)=
\left\{\begin{array}{lll} e^{i k x}+\rho_s^r(E)e^{-i k x} \\
A^r_s(E) e^{i q x}+B^r_s(E)e^{-i q x} \\
\sigma^r_s(E)e^{i k x}
\end{array}\right.
\hspace{0.2cm} , \hspace{0.2cm}
\psi^{l(s)}(x,E)=
\left\{\begin{array}{lll} \sigma^l_s(E)e^{-i k x} \\
A^l_s(E) e^{i q x}+B^l_s(E) e^{-i q x} \\
e^{-i k x}+\rho^l_s(E)e^{i k x} \end{array}\right.   \begin{array}{lll} , \ \ x\in \text{II} \\ , \ \ x\in \text{I} \\
, \ \ x\in \text{III} \end{array}
\end{equation*}
where $k=\sqrt{E^{(s)}-\alpha^{2}}$ and $q=\sqrt{E^{(0)}}$\ {\footnote{There would be different labels in each sector: $\psi^r_0$, $\psi^r_1$, $\psi^l_0$, $\psi^l_1$, $E^{(0)}$, $E^{(1)}$,
and so on.}}.
As in the non-SUSY case, the system of four linear equations in the four unknowns $\sigma$, $A$, $B$, $\rho$ set by the matching conditions is easy to solve, and we find, e.g., in the bosonic sector the following scattering matrix:
\begin{equation*}
S=\left( \begin{array}
[c]{cc}%
-\frac{4e^{-2ia(k-q)}kq}{e^{4iaq}(k-q+i\alpha)^{2}-(k+q+i\alpha)^{2}}&
\frac{e^{-2iak}[(e^{4iaq}-1)k^{2}-e^{4iaq}(q-i\alpha)^{2}+(q+i\alpha)^{2}]}{e^{4iaq}(k-q+i\alpha)^{2}-(k+q+i\alpha)^{2}}\\
\frac{e^{-2iak}[(e^{4iaq}-1)k^{2}-e^{4iaq}(q-i\alpha)^{2}+(q+i\alpha)^{2}]}{e^{4iaq}(k-q+i\alpha)^{2}-(k+q+i\alpha)^{2}}& -\frac{4e^{-2ia(k-q)}kq}{e^{4iaq}(k-q+i\alpha)^{2}-(k+q+i\alpha)^{2}}
\end{array} \right)
 \label{Smddc0}
\end{equation*}
encoding the scattering amplitudes{\footnote{$V_0$ and $V_1$ are even in the $x\to -x$ exchange and give rise to time reversal invariant Hamiltonians.}} $\sigma_{0}^r
=\sigma_{0}^l=\sigma_0$ and $\rho_{0}^r=\rho_{0}^l=\rho_0$.
\\
The eigenvalues of the unitary $S$ matrix
\begin{equation*}
\lambda_{0+}=e^{2i\delta_{0+}}=\frac{e^{-2 i a k} [(k-i \alpha ) \cos qa+i q \sin qa]}{(k+i \alpha ) \cos qa-i q \sin qa}
\, \, \, ; \, \, \, 
\lambda_{0-}=e^{2i\delta_{0-}}=-\frac{e^{-2 i a k} [(k-i \alpha ) \sin qa-i q \cos qa]}{(k+i \alpha ) \sin qa+i q \cos qa}
\end{equation*}
provide the total phase shift in the bosonic sector:
\begin{equation*}
\delta_0=\delta_{0+}+\delta_{0-}=\frac{1}{2 i}  \ln \left[\frac{e^{-4 i a k} (\alpha +i k) \left(q \cos 2
   qa+i k \sin 2 qa\right)}{(k+i \alpha
   ) \left(k \sin 2 qa+i q \cos 2 qa\right)}\right]
\end{equation*}
where $q=\sqrt{E^{(0)}}=\sqrt{k^2+\alpha ^2}$ for the scattering solutions. Thus, the spectral density for the bosonic scattering waves with periodic boundary conditions on a line of very long length $L$ becomes
\begin{equation*}
\varrho_0(k)=\frac{L}{2\pi}+\frac{1}{2\pi}\frac{d\delta_0}{dk}\, \, , \, \, \frac{d\delta_0}{dk}=\frac{\alpha  q \left[\alpha ^2-2 a \alpha ^3+k^2 (2-2 a \alpha )\right]+\alpha ^2 q^2 \sin 4 qa+\alpha ^2 q \left(\alpha -2 a q^2\right) \cos 4 qa}{q^3 \left(\alpha ^2+\alpha ^2 \cos 4 q a+2 k^2\right)}
\end{equation*}
The scattering amplitudes in the fermionic sector driven by $V_1$ are obtained by replacement of $\alpha$ by $-\alpha$.
The analogous identities to \eqref{igual} between the scattering amplitudes in the fermionic and bosonic sectors of the SUSY two-$\delta$ problem are:
\begin{equation*}
\rho_{1}^r=\frac{-i k+\alpha}{i k+\alpha} \ \rho_{0}^r \quad , \, \, \sigma_{1}^r=\frac{i k-\alpha}{i k+\alpha} \ \sigma_{0}^r \label{igual1}
\end{equation*}
meaning that the scattering wave functions in the sectors with different Fermi number are paired through the supercharges. It is easy to check that the bosonic and fermionic right-to-left waves satisfy the same relations changing $\alpha$ by $-\alpha$. Even though $\rho_0\neq\rho_1$ and $\sigma_0\neq\sigma_1$, the reflection and transmission coefficients are equal in the bosonic and fermionic sectors: $\left|\rho_{0}^r\right|^2=\left|\rho_{1}^r\right|^2$, $\left|\sigma_{0}^r\right|^2=\left|\sigma_{1}^r\right|^2$.

\subsubsection{Bound/anti-bound states}
The bound states have energy $0<E<\alpha^2$ and are solutions of the form given by equation  (53) in reference \cite{Munoz-Castaneda-prd2013} with $iq$  instead of $\kappa$ in the zone I ($-a<x<a$). The values of $\kappa$ providing bound/anti-bound solutions are the positive/negative imaginary part of the purely imaginary poles of the transmission amplitudes. Starting with the bosonic sector the purely imaginary poles of $\sigma_0$ are the solutions of the transcendent equation
\begin{eqnarray}
&& e^{4iaq}(i(\kappa+\alpha)-q)^{2}=(i(\kappa+\alpha)+q)^{2} \quad , \quad \kappa=\sqrt{\alpha^{2}-E^{(0)}} \nonumber\\
&& e^{4iaq}= \left[\frac{q^2-(\alpha+\kappa)^2+2iq(\kappa+\alpha)}{q^2+(\kappa+\alpha)^2}\right]^2 \, \, \, \equiv \, \, \,
{\rm tan}2aq = \frac{2q(\alpha+\kappa)}{q^2-(\kappa+\alpha)^2} \label{bssc} \quad .
\end{eqnarray}

Because ${\rm tan}(2z)=\frac{2}{{\rm cot}z-{\rm tan}z}$ this equation (\ref{bssc}) is equivalent to ${\rm cot}qa-{\rm tan}qa=\frac{q}{\alpha +\kappa}-\frac{\alpha+\kappa}{q}$. The bound state spectral condition (\ref{bssc}) decomposes into two kinds of alternative (simpler) spectral conditions:
\begin{equation}
{\rm (e)} \, \, \, \sin qa=\frac{\kappa+\alpha}{q}\cos qa \hspace{1cm} , \hspace{1cm} {\rm (o)} \, \, \, \cos qa=-\frac{\kappa+\alpha}{q}\sin qa\quad . \label{eepddcr}
\end{equation}
This factorization is due to the fact that there is parity invariance when $\alpha=\beta$ and the wave eigen-functions are either even or odd functions of $x$.

\begin{figure}[htbp]
\begin{center}
\begin{tabular}{|p{6.5cm}|p{6.5cm}|}
\hline & \\
\includegraphics[scale=0.475]{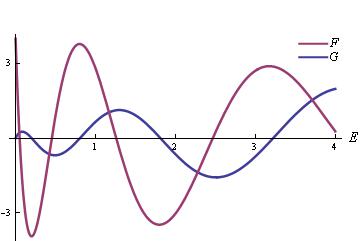} & \includegraphics[scale=0.475]{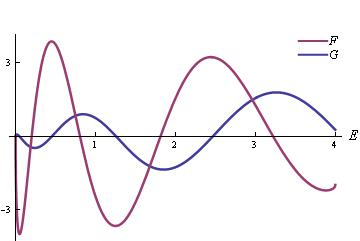}  \\
 \hline &\\ \includegraphics[scale=0.475]{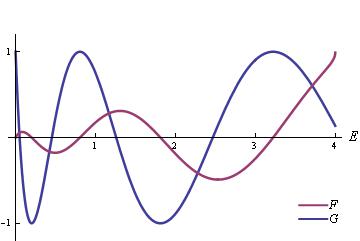} & \includegraphics[scale=0.475]{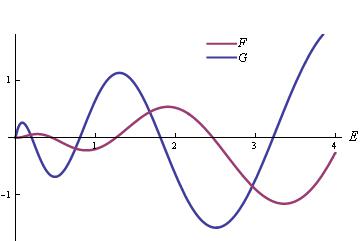}  \\
\hline
 \end{tabular} 
 \caption{Graphics of the curves in the two members of the spectral equations \eqref{eepddcr}. The particular values $a=7$ and $\alpha=2$
have been selected for the plots. In the left column the curves describing the two members of \eqref{eepddcr}(e) for $\alpha=2$ (upper box, $F=\left(\sqrt{4-E}+2\right) \cos 7 \sqrt{E}$, $G=\sqrt{E} \sin 7 \sqrt{E}$, henceforth identifying the bosonic even bound states) and  \eqref{eepddcr}(o) for $\alpha=-2$ (lower box, $F=-\left(\sqrt{4-E}-2\right) \sin 7\sqrt{E}$, $G=\sqrt{E}\cos 7 \sqrt{E}$, henceforth corresponding to the odd fermionic bound states) are depicted as functions of $E$.
Therefore, the values  of $E$ for which the curves intersect are the even bound state eigenvalues of $\hat{H}_0$ and the odd bound state eigenvalues of $\hat{H}_1$. Because the cuts happen between $0$ and $4$ the corresponding values of $\kappa_B^{(0)}$ and  $\kappa_B^{(1)}$
are between $0$ and $2$. Observe that in all the cases $\kappa_B^{(0)}=\kappa_B^{(1)}>0$ as it should be in a supersymmetric spectrum.
In the right column we plot the curves in \eqref{eepddcr}(e) for $\alpha=-2$ (lower box, $F=\left(\sqrt{4-E}-2\right) \cos 7 \sqrt{E}$, $G=\sqrt{E} \sin 7 \sqrt{E}$, henceforth describing the fermionc even bound states) and \eqref{eepddcr}(o) for $\alpha=2$ (upper box, $F=\left(-\sqrt{4-E}-2\right) \sin 7 \sqrt{E}$, $G=\sqrt{E} \cos 7 \sqrt{E}$, henceforth identifying the bosonic odd bound states). We check that the only zero mode is fermionic, see the intersection of the two curves at $E_+^{(1)}=0$ in the lower right box. The choice $\alpha=2>0$ means that the two-$\delta$ potential in the $\hat{H}_0$ Hamiltonian is repulsive. The three lower positive energy eigenvalues read from the intersection points of the curves are: $E_{1+}^{0)}=E_{1-}^{(1)}=0.0469$, $E_{2-}^{0)}=E_{2+}^{(1)}=0.187$, $E_{3+}^{0)}=E_{3-}^{(1)}=0.421$.} \label{ddc}
\end{center}
\end{figure}

\begin{itemize}

\item \underline{Even bound states}. The transcendent equation that gives the eigenvalues of the even bound states is given by (\ref{eepddcr})(e) with $q=\sqrt{E^{(0)}_+}$ and $\kappa=\sqrt{\alpha^2-E^{(0)}_+}$.  The positive solutions $\kappa_B=\sqrt{\alpha^2-E^{(0)}_+}>0$ of equation (\ref{eepddcr})(e) are the even bound states of $\hat{H}_0$ (bosonic):
    \begin{equation*}
    \psi^{(0)}_{B_+ II}(x)=Ae^{\kappa_{B_+} x} \, \, , \, \, \psi^{(0)}_{B_+ I}(x)=A\left(\frac{e^{-\kappa_{B_+} a}}{{\rm cos}q_{B_+}a}\right){\rm cos}q_{B_+}x \, \, , \, \, \psi^{(0)}_{B_+ III}(x)=Ae^{-\kappa_{B_+} x} \label{bebs}
    \end{equation*}

\item \underline{Odd bound states}. With similar notation ($q=\sqrt{E^{(0)}_-}$ and $\kappa=\sqrt{\alpha^2-E^{(0)}_-}$) the energies $E^{(0)}_-$ of the odd bound states are given by the transcendent equation (\ref{eepddcr})(o). The positive solutions $\kappa_B=\sqrt{\alpha^2-E^{(0)}_-}>0$ of equation (\ref{eepddcr})(o) are the odd bound states of $\hat{H}_0$ (bosonic):
    \begin{equation*}
    \psi^{(0)}_{B_- II}(x)=Ae^{\kappa_{B_-} x} \, \, , \, \, \psi^{(0)}_{B_-I}(x)=A\left(\frac{e^{-\kappa_{B_-} a}}{{\rm sin}q_{B_-}a}\right){\rm sin}q_{B_-}x \, \, , \, \, \psi^{(0)}_{B_- III}(x)=-Ae^{-\kappa_{B_-} x} \label{bobs}
    \end{equation*}
\end{itemize}
The spectral equations can only be solved by graphic methods (see Figure \ref{ddc}).

\par
The positive values of $\kappa$ where the curves in the
left and right members of (\ref{eepddcr})(e) and (\ref{eepddcr})(o) intersect provide the bosonic even
and odd bound states respectively. It is clear that the even spectral condition admits a solution $\kappa_0=-\alpha$, $q_0=0=E^{(0)}_{0+}$ that is a bona fide ground state if $\alpha<0$. This (always even) bound state is the bosonic ground state. Regardless the sign of $\alpha$, the odd spectral condition presents no zero energy bound state because the cotangent at zero angle is infinity. Nevertheless, the first intersection in (\ref{eepddcr})(o) is the next bound state in energy $E^{(0)}_{1-}$, there is then an even bound state and so on: $0=E^{(0)}_{0+}<E^{(0)}_{1-}<E^{(0)}_{2+}<\cdots <\alpha^2$.

Regarding the positive fermionic bound states of $\hat{H}_1$ everything is the same if one changes
$\alpha$ by $-\alpha$. Because
\begin{eqnarray*}
&& {\rm tan}\sqrt{E^{(1)}}a=\frac{-\alpha+\sqrt{\alpha^2-E^{(1)}}}{\sqrt{E^{(1)}}} \, \, \, \equiv \, \, \,
{\rm cot}\sqrt{E^{(1)}}a=-\frac{\alpha+\sqrt{\alpha^2-E^{(1)}}}{\sqrt{E^{(1)}}} \\ && {\rm cot}\sqrt{E^{(1)}}a=-\frac{-\alpha+\sqrt{\alpha^2-E^{(1)}}}{\sqrt{E^{(1)}}} \, \, \, \equiv \, \, \,
{\rm tan}\sqrt{E^{(1)}}a=\frac{\alpha+\sqrt{\alpha^2-E^{(1)}}}{\sqrt{E^{(1)}}}
\end{eqnarray*}
$E^{(1)}_{(n+1)+}=E^{(0)}_{(n+1)-}$, $E^{(1)}_{(n+2)-}=E^{(0)}_{(n+2)+}$, $n=0,1,2, \cdots$ and the fermionic and bosonic bound states are paired by supersymmetry. Note that the supercharges necessarily change the parity:
\[
\hat{Q}^\dagger\left(\begin{array}{c} \psi_B^{(0)}(x) \\ 0 \end{array}\right)=\left(\begin{array}{c} 0 \\\psi_B^{(1)}(x)\end{array}\right) \quad , \quad \hat{Q}\left(\begin{array}{c} 0 \\ \psi_B^{(1)}(x) \end{array}\right)=\left(\begin{array}{c} \psi_B^{(0)}(x) \\ 0 \end{array}\right) \quad .
\]
$\hat{H}_0$ and $\hat{H}_1$, however, are not completely isospectral. The zero energy
ground state is either bosonic, if $\alpha<0$, or fermionic, when $\alpha$ is positive. Note that $\kappa_0^{(0)}=-\alpha \,\equiv\, E_0^{(0)}=0$ whereas $\kappa_0^{(1)}=\alpha \,\equiv\, E_0^{(1)}=0$.
Thus, one of the two zero modes is an anti-bound state because either $\kappa_0^{(0)}$ or $\kappa_0^{(1)}$
is negative. The Witten index is one{\footnote{In fact, there is an infinite contribution to the Witten index due to the continuous spectrum. Conveniently regularized, the difference
between the bosonic and fermionic spectral densities induces the result: $I_W=2$.}} and supersymmetry is not spontaneously
broken. The ground states (zero energy) belong simultaneously to the kernels of $\hat{Q}$ and $\hat{Q}^\dagger$ being the exponential of either minus (bosonic) or plus (fermionic) the superpotential. The key
fact is that the ground states are singlets, not paired states, of the supersymmetry algebra.
\\
We choose now $\alpha>0$ to describe explicitly stationary wave functions. The unique zero energy $q_0=E^{(1)}=0$, $\kappa_0=\alpha$ (normalized) ground state is fermionic and even:
$$
\psi_{0}^{(1)}(x)=N{\rm exp}\left[-\frac{\alpha}{2}|x-a|-\frac{\alpha}{2}|x+a|\right]=\sqrt{\frac{\alpha}{1+2\alpha a}}\left\{ \begin{array}{lll}
e^{\alpha(x+a)} & \, , \, \, \, \, x\in \text{II}\\
1 & \, , \, \, \, \, x\in \text{I}\\
e^{-\alpha(x-a)} & \, , \, \, \, \, x\in \text{III}
\end{array} \right. \, \, \, ,
$$
\\
The remaining non-zero energies of the bosonic and fermionic bound states $E^{(0)}=E^{(1)}\neq0$ cannot be given analytically because they are graphically determined as the intersections of the curves in the two members of the transcendent equations (\ref{eepddcr})(e)-(o). From the $n$-th intersection values $\kappa_n=\sqrt{\alpha^2-E_n^{(0)}}$, $q_n=\sqrt{E^{(0)}}$, one obtains:
\begin{equation}
\psi_{n+}^{(0)}(x)= \sqrt{\frac{\kappa_n}{1+\kappa_n a+\frac{\alpha}{q_n}\sin q_na\cos q_na}} \left\{ \begin{array}{lll}
\cos q_na \ e^{\kappa_n(x+a)} & \, , \, \, \, \, x\in \text{II}\\
\cos q_nx & \, , \, \, \, \, x\in \text{I}\\
\cos q_na \ e^{-\kappa_n(x-a)} & \, , \, \, \, \, x\in \text{III}\\
\end{array} \right. \label{wfpddcea}
\end{equation}
\begin{equation}
\psi_{n-}^{(0)}(x)= \sqrt{\frac{\kappa_n}{1+\kappa_n a+\frac{\alpha}{q_n}\sin q_na\cos q_na}} \left\{ \begin{array}{lll}
-\sin q_na \ e^{\kappa_n(x+a)} & \, , \, \, \, \, x\in \text{II}\\
\sin q_nx & \, , \, \, \, \, x\in \text{I}\\
\sin q_na \ e^{-\kappa_n(x-a)} & \, , \, \, \, \, x\in \text{III}\\
\end{array} \right. \label{wfpddcer}
\end{equation}

The underlying reason of the need of one square well to be added to the two $\delta$s in order to achieve a supersymmetric quantum mechanical system can be explained as follows. The analogous factorization of the spectral condition into the two (\ref{eepddcr})(e)-(o) equations in the non-SUSY two $\delta$s problem is:
\begin{equation}
(e) \, \, e^{-2 a \sqrt{|E|}}=-1-\frac{2\sqrt{|E|}}{\alpha} \quad , \quad ({\rm o}) \, \, e^{-2 a \sqrt{|E|}}=1+\frac{2\sqrt{|E|}}{\alpha}\label{2dbseo} \quad .
\end{equation}
If $|E_0|$ and $|E_1|$ are the solutions of (\ref{2dbseo})(e) and (\ref{2dbseo})(o) respectively then
$|E_1|<|E_0|$, which means that the even bound state has less energy than the odd one because both bound states are of negative energy. Therefore, the role of the well is to push the deepest bound state energy to zero.

\subsection{$\mathcal{N}=2$ supersymmetric two Dirac-$\delta$ Hamiltonian: two $\delta$s of different strength}
If the two $\delta$s are of different strengths we choose the superpotential:
\begin{equation*}
W(x)=\frac{\alpha}{2}\left|x+a\right|+\frac{\beta}{2}\left|x-a\right|-\frac{1}{2}(\alpha-\beta)x ; \, \, \, \, v_-=-\alpha , \, \, \, \, v_+=\beta
\end{equation*}
the reason for this selection will be clear later. The intertwined SUSY partner Hamiltonians are
\begin{align*}
\hat{H}_s=&-\frac{d^{2}}{dx^{2}}+(-1)^s\alpha\delta(x+a)+(-1)^s\beta\delta(x-a)+\frac{\alpha\beta}{2}\epsilon(x+a)\epsilon(x-a)\nonumber\\
& -(\alpha-\beta)\left[\frac{\alpha}{2}\varepsilon(x+a)
+\frac{\beta}{2}\varepsilon(x-a)\right]+\frac{\alpha^2}{2}+\frac{\beta^2}{2}-\frac{\alpha\beta}{2}
\end{align*}
where $s=0,1$ is the Fermi number. The potential energies $V_0(x)$ and $V_1(x)$ are composed of the two $\delta$s plus one quasi-square well:
\begin{eqnarray}
&& V_s(x)=(-1)^s\alpha\delta(x+a)+(-1)^s\beta\delta(x-a)+\left\{\begin{array}{lll}
\alpha^2 \\
0 \\
\beta^2
\end{array}\right. \begin{array}{lll} , \ \ x\in \text{II} \\ , \ \ x\in \text{I} \\
, \ \ x\in \text{III} \end{array} ,\quad s=0,1.\nonumber
\end{eqnarray}
The momenta of the wave functions in the three zones are different: Zone II , $k=\sqrt{E^{(s)}-\alpha^{2}}$. Zone I, $q=\sqrt{E^{(s)}}$. Zone III, $p=\sqrt{E^{(s)}-\beta^{2}}$. We assume, without loss of generality, that $\alpha^2<\beta^2$. Note the similarity with the $\delta$/step potential. There are three ranges of energies: $E^{(s)}>\beta^2$, $\alpha^2<E^{(s)}<\beta^2$ and $0\leq E^{(s)}<\alpha^2$ .
\paragraph{Double degenerate scattering waves.}
If $E^{(s)}>\beta^2$ there are left-to-right and right-to-left scattering wave solutions of the bosonic ($s=0$) and fermionic ($s=1$) Schr$\ddot{\rm o}$dinger equations:

\begin{equation*}
\psi^{r(s)} (x,E)=
\left\{\begin{array}{lll} e^{i k x}+\rho_s ^r(E)e^{-i k x} \\
A^r_s(E) e^{i q x}+B^r_s(E )e^{-i q x} \\
\sigma^r_s(E)e^{i p x}
\end{array}\right.
\hspace{0.2cm} , \hspace{0.2cm}
\psi^{l(s)}(x,E)=
\left\{\begin{array}{lll} \sigma^l_s(E)e^{-i k x} \\
A^l_s(E) e^{i q x}+B^l_s(E ) e^{-i q x} \\
e^{-i p x}+\rho^l_s(E)e^{i p x} \end{array}\right.   \begin{array}{lll} , \ \ x\in \text{II} \\
, \ \ x\in \text{I} \\ , \ \ x\in \text{III} \end{array}
\end{equation*}
Solving the linear system, which arise from imposing the matching conditions on the scattering waves, we find the scattering amplitudes:
\begin{eqnarray*}
&& \rho^r_0(E)=\frac{e^{-2 i a k} \left[e^{4 i a q} (k+q-i \alpha ) (-p+q-i \beta )+(k-q-i \alpha ) (p+q+i \beta)\right]}{\Delta_0(E,\alpha,\beta,a)} \\
&& 
\sigma^r_0(E)=\frac{4 k q e^{-i a (k+p-2 q)}}{\Delta_0(E,\alpha,\beta,a)} \quad , \quad
\Delta_0(E,\alpha,\beta,a)=(k+q+i \alpha) (p+q+i \beta)-e^{4 i a q} (k-q+i \alpha) (p-q+i \beta) \, ,
\end{eqnarray*}
whereas the left incoming scattering amplitudes are obtained through the exchange $\alpha\leftrightarrow \beta$. We stress that the zone III momentum is $p$ as the only difference with respect to the symmetric case where it is identical to the zone I momentum $k$.

Since the fermionic scattering amplitudes are obtained from the bosonic ones by changing $\alpha$ by $-\alpha$ and $\beta$ by $-\beta$, the bosonic scattering coefficients (of $\hat{H}_0$) are identical to the fermionic scattering coefficients (of $\hat{H}_1$). In fact, the following identities between the bosonic and fermionic scattering amplitudes hold:
\begin{equation*}
\rho^r_1(E)=\frac{-i k+\alpha}{i k+\alpha} \rho^r_0(E) \quad , \quad \sigma^r_1(E)=\frac{i p-\beta}{i k+\alpha}  \sigma^r_0(E)\quad ,
\end{equation*}
and the corresponding ones between the right-to-left amplitudes.  Accordingly, the transmission and reflection bosonic and fermionic probabilities are equal: $\left|\rho_{0}^r\right|^2=\left|\rho_{1}^r\right|^2$ and $\left|\sigma_{0}^r\right|^2=\left|\sigma_{1}^r\right|^2$, $\left|\rho_{0}^l\right|^2=\left|\rho_{1}^l\right|^2$ and $\left|\sigma_{0}^l\right|^2=\left|\sigma_{1}^l\right|^2$. Also, we have: $\bigtriangleup_1(E,\alpha,\beta,a)=\bigtriangleup_0(E,-\alpha,-\beta,a)$.
\paragraph{Only left-to-right scattering waves.}
When $\alpha^2<E^{(s)}<\beta^2$ the momentum in zone III becomes purely imaginary: $p=i\sqrt{\beta^2-E^{(s)}}$. Hence there are no incoming right-to-left waves but the left-to-right waves are partly reflected and the transmitted waves decay exponentially.

\paragraph{Bound/anti-bound states.}
For energies such that $0\leq E^{(s)}\leq\alpha^2$, corresponding to purely imaginary momenta
 $k^{(s)}=i\kappa^{(s)}=i\sqrt{\alpha^{2}-E^{(s)}}$ and $p^{(s)}=i\mathcal{\pi}^{(s)}=i\sqrt{\beta^{2}-E^{(s)}}$ (positive imaginary part: $\kappa^{(s)}
,\mathcal{\pi}^{(s)}>0$),
the poles of the scattering amplitudes occur at the bound state energies. The poles are obviously the zeroes of $\bigtriangleup_s(E,\alpha,\beta,a)$. Henceforth in the bosonic sector,
solving for $E$ the spectral equation
\begin{equation}
\Delta_0(E^{(0)},\alpha,\beta,a)=0 \, \, \equiv \, \, \frac{q(\kappa^{(0)}+\mathcal{\pi}^{(0)}+\alpha+\beta)}{q^{2}-(\kappa^{(0)}+\alpha)(\mathcal{\pi}^{(0)}+\beta)}\cos 2qa=\sin 2qa \label{eepddccr}
\end{equation}
we obtain the bound state energies of $\hat{H}_0$. With the required changes, in the fermionic sector the solutions in $E$ of the spectral equation
\begin{equation}
\Delta_1(E^{(1)},\alpha,\beta,a)=0 \, \, \equiv \, \, \frac{q(\kappa^{(1)}+\mathcal{\pi}^{(1)}-\alpha-\beta)}{q^{2}-(\kappa^{(1)}-\alpha)(\mathcal{\pi}^{(1)}-\beta)}\cos 2qa=\sin 2qa. \label{eepddcca}
\end{equation}
are the bound state eigenvalues.
\par
\begin{figure}[htdp]
\begin{center}

\begin{tabular}{|p{6.5cm}|p{6.5cm}|}
\hline & \\
\includegraphics[scale=0.475]{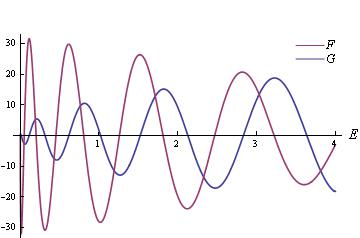} & \includegraphics[scale=0.475]{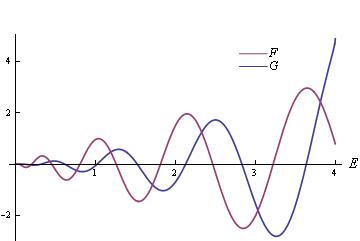}  \\
 \hline
 \end{tabular} 
 \caption{Graphics of the curves in the spectral equations \eqref{eepddccr} (left)  and  \eqref{eepddcca} (right) for  $a=7$, $\alpha=2$ and $\beta=4$. The intersection points give the
bound state eigenvalues for $\hat{H}_0$ (left, $F=\left[x-\left(\sqrt{4-x}+2\right)
   \left(\sqrt{16-x}+4\right)\right] \sin 14 \sqrt{x}$, $G=\left(\sqrt{4-x}+\sqrt{16-x}+6\right) \sqrt{x} \cos 14
   \sqrt{x}$) and $\hat{H}_1$ (right, $F=\left[x-\left(\sqrt{4-x}-2\right)
   \left(\sqrt{16-x}-4\right)\right] \sin 14
   \sqrt{x}$, $G=\left(\sqrt{4-x}+\sqrt{16-x}-6\right) \sqrt{x} \cos 14
   \sqrt{x}$). We remark that $E_B^{(0)}=E_B^{(1)}=x$.}\label{ddcc}
\end{center}
\end{figure}

Since
\begin{eqnarray}
&&\frac{q(\sqrt{\alpha^2-E}+\sqrt{\beta^2-E}\pm(\alpha+\beta))}{q^{2}-(\sqrt{\alpha^2-E}\pm\alpha)(\sqrt{\beta^2-E}\pm\beta)}
=-\frac{\beta\sqrt{\alpha^2-E}-\alpha\sqrt{\beta^2-E}}{(\alpha-\beta)q}
\nonumber
\end{eqnarray}
for arbitrary $\alpha$ and $\beta$, the solutions for $E^{(0)}$ and $E^{(1)}$ of the transcendent equations (\ref{eepddccr}) and (\ref{eepddcca})  are the same.
 As expected, there is pairing between the bound states of the intertwined operators $\hat{H}_0(x)$ and $\hat{H}_1(x)$, see Figure \ref{ddcc}.
\par
Both $\hat{H}_0$ and $\hat{H}_1$ have a zero energy eigenstate, as one can check by taking the $E\to 0$ limit of the spectral equations. Nevertheless, for the particular choice where $\alpha$ and $\beta$ are positive, the wave eigenfunction in the kernel of $\hat{H}_0$ is not normalizable because $W\rightarrow+\infty$ in the  $x\rightarrow\pm\infty$ limits. The zero energy wave function of  $\hat{H}_1$, however, is normalizable: $\psi_0^{(1)}(x)\propto e^{-W(x)}$ tends to zero at both ends of the real line.
\\
In fact, the fermionic zero mode wave function properly normalized is:\footnote{Recall the redefinitions: $\frac{2m}{\hbar^{2}}E\rightarrow E$ and $\frac{2m}{\hbar^{2}}V(x)\rightarrow V(x)$}
$$
\psi_{0}^{(1)}= \sqrt{\frac{2\alpha}{\alpha+4a\alpha\beta+\beta}}\left\{ \begin{array}{lll}
e^{\alpha(x+a)} & \, , \, \, \, \, x\in \text{II}\\
1 & \, , \, \, \, \, x\in \text{I}\\
e^{-\beta(x-a)} & \, , \, \, \, \, x\in \text{III}
\end{array} \right. \quad .
$$

For energies in the range $0<E^{(0)}<\alpha^2$ we label as $\kappa_n$, $\mathcal{\pi}_n$, $q_n$ the momenta in the three zones corresponding to the bound state energies that solve the transcendent bosonic spectral equation. The normalized bosonic wave functions are:
\begin{equation*}
\psi_n^{(0)}(x)= \sqrt{\frac{\kappa_n}{\frac{r_n}{2}+\kappa_n(m_n+\frac{1}{2\mathcal{\pi}_n})}} \left\{\begin{array}{llll}
\vspace{-0.3cm}\\
\ \left(\cos2q_na+\frac{\mathcal{\pi}_n-\beta}{q_n}\sin2q_n a\right) \ e^{\kappa_n(x+a)} & \, , \, \, \, \, x\in \text{II} \vspace{0.1cm}\\
\ \cos q_n(x-a)-\frac{\mathcal{\pi}_n-\beta}{q_n}\sin q_n(x-a) & \, , \, \, \, \, x\in \text{I}\vspace{0.1cm}\\
\ e^{-\mathcal{\pi}_n (x-a)} & \, , \, \, \, \, x\in \text{III}\vspace{0.1cm}\\
\end{array} \right.  \label{wfpddccr}
\end{equation*}
where $r_n$ and $m_n$ are defined as:
$$
r_n=\left(\cos2q_n a+\frac{\mathcal{\pi}_n-\beta}{q_n}\sin2q_n a\right)^{2},
$$
$$
m_n=\frac{2q_n a\left[q_n^{2}+(\mathcal{\pi}_n-\beta)^{2}\right]+\sin2q_n a\left\{\left[q_n^{2}-(\mathcal{\pi}_n-\beta)^{2}\right]\cos2q_n a
+2(\mathcal{\pi}_n-\beta)q_n\sin2q_n a\right\}}{2q_n^{3}}.
$$
Again, the fermionic bound state wave functions are obtained by applying the supercharge to  the bosonic ones: $\psi_n^{(1)}(x)=\hat{Q}^\dagger\psi_n^{(0)}(x)$.

\section{The $\mathcal{N}=2$ supersymmetric triple Dirac-$\delta$ potential}\label{s6} 
In this section we will study in detail the SUSY quantum mechanical system of three Dirac $\delta$s. This system provides the necessary information to understand any SUSY configuration with a finite number of Dirac $\delta$s and also establishes  basis to deal with a system of infinite $\delta$s. From a quantum field theoretical point of view the system of three supersymmetric Dirac $\delta$s can be interpreted as a SUSY version of the piston geometries modelled by point interactions, see refs. \cite{fucci-spp2011,fucci-ijmp2012}, as well as supersymmetric Kaluza-Klein field theories when these theories are interpreted as field theories defined over piston geometries as it was done in ref. \cite{kirsten-prd09}).

\subsection{$\mathcal{N}=2$ supersymmetric three Dirac $\delta$ Hamiltonian: three $\delta$s of different strengths}
The supersymmetric quantum mechanics of three $\delta$ configurations of different strength is determined from the superpotential
\begin{equation*}
W(x)=\frac{\alpha}{2}\left|x+a\right|+\frac{\mu}{2}\left|x\right|+\frac{\beta}{2}\left|x-a\right|-\frac{1}{2}(\alpha-\beta)x, \label{wtd}
\end{equation*}
where $\alpha\neq \beta$, $\alpha\neq \mu$, $\beta\neq \mu$ are three different positive constants. It is clear that
\begin{eqnarray*}
&& W^\prime(x)=\frac{\alpha}{2} \varepsilon(x+a)+\frac{\mu}{2} \varepsilon(x)+\frac{\beta}{2} \varepsilon(x-a)-\frac{\alpha-\beta}{2} \quad , \quad v_+=\beta+\frac{\mu}{2} \, \, \, , \, \, \,
v_-=-\alpha-\frac{\mu}{2}\\ && \lim_{x\to 0_\pm} W^\prime(x)=\pm \frac{\mu}{2} \qquad , \qquad \pm W^{\prime\prime}(x)=\pm \left[\alpha\delta(x+a)+\mu\delta(x)+\beta\delta(x-a)\right]
\end{eqnarray*}
The bosonic scalar Hamiltonian is the Schr$\ddot{\rm o}$dinger operator for three repulsive $\delta$s plus one quasi-square well whereas the fermionic Hamiltonian is obtained by replacing the repulsive $\delta$-walls by attractive $\delta$-wells:
\begin{eqnarray}
&&\hspace{1.11cm}\hat{H}_0=-\frac{d^{2}}{dx^{2}}+V_0(x) \qquad , \qquad \hat{H}_1=-\frac{d^{2}}{dx^{2}}+V_1(x) \nonumber\\
&& V_s(x)=(-1)^s\alpha\delta(x+a)+(-1)^s\mu\delta(x)+(-1)^s\beta(x-a)+\left\{\begin{array}{lll} (\alpha+\frac{\mu}{2})^2 \\ \frac{\mu^2}{4} \\
(\beta+\frac{\mu}{2})^2 \end{array}\right. \begin{array}{lll} , \ \ x\in \text{III} \\ , \ \ x\in \text{I}, \text{II} \\
, \ \ x\in \text{IV} \end{array} \nonumber \quad .
\end{eqnarray}
As in the case of two different $\delta$s we introduce a quasi-square well, in this case different from zero also in the middle region, to make 
the problem compatible with supersymmetry. There are four zones separated by the three $\delta$ walls/wells and only three different momenta: zone III , $k=\sqrt{E^{(s)}-(\alpha+\frac{\mu}{2})^{2}}$; Zones I, and II: $q=\sqrt{E^{(s)}-\frac{\mu^2}{4}}$; zone IV, $p=\sqrt{E^{(s)}-(\beta+\frac{\mu}{2})^{2}}$. We assume, without loss of generality, that $\alpha^2<\beta^2$.
\subsubsection{Double degenerate scattering waves}
If $E^{(s)}>(\beta+\frac{\mu}{2})^2$ there are right-going and left-going scattering wave solutions of the bosonic and fermionic Schr$\ddot{\rm o}$dinger equations. The solution of the linear system of six equations in six unknowns arising from the matching conditions at the points $x=\pm a$ and $x=0$ on the scattering waves, e.g., for the right-movers in the bosonic sector, gives the scattering amplitudes:
\begin{eqnarray*}
&&\hspace{-0.3cm}\sigma_{0}^r(E)=
\frac{8 k q^2 e^{-i a (k-3 q)}}{\Delta ^r_0} \\
&&\hspace{-0.8cm}\rho_{0}^r(E)=e^{-2 i a k} \frac{k+q-i \alpha}{k-q+i \alpha}-\frac{4 k q e^{i a (p+q-2k)}}{(k-q+i \alpha)\Delta ^r_0}\left[\mu  e^{2
   i a q} (\beta -i p+i q)+(2 q+i \mu ) (i \beta +p+q)\right] \\
&&\hspace{-0.8cm}
\Delta_{0}^r(E,\alpha,\mu,\beta,a)=2 \mu  e^{i a (p+3 q)} \left[(\alpha -i k) (p+i \beta )+ i q^2\right]+\\ && +e^{i a (p+q)}(2 q+i \mu ) (k+q+i \alpha)
(p+q+i \beta)+e^{i a (p+5 q)}(2 q-i \mu )
   (k-q+i \alpha) (-p+q-i \beta ) \quad .
\end{eqnarray*}
The scattering amplitudes in the fermionic sector for the right-movers are then determined from supersymmetry:
\begin{equation*}
\sigma_{1}^r(E)=\frac{i p-\beta-\frac{\mu }{2}}{i k+\alpha+\frac{\mu }{2}} \sigma_{0}^r(E) \qquad , \qquad \rho_{1}^r(E)=\frac{-i k+\alpha+\frac{\mu }{2}}{i k+\alpha+\frac{\mu }{2}} \ \rho_{0}^r(E)
\end{equation*}
Similar formulas describe the scattering amplitudes of the waves incoming from the right.
\subsubsection{Bound state wave functions}
For energies such that $(\alpha+\frac{\mu}{2})^2<E^{(s)}<(\beta+\frac{\mu}{2})^2$ there are no left-going incoming scattering waves and only exponentially decaying right-going transmitted waves.
If $\frac{\mu^2}{4}<E^{(s)}<(\alpha+\frac{\mu}{2})^2$, however, the two asymptotic momenta become purely imaginary: $k^{(s)}=i\kappa^{(s)}=i\sqrt{\left(\alpha+\frac{\mu}{2}\right)^2-E^{(s)}}$, $p^{(s)}=i\mathcal{\pi}^{(s)}=i\sqrt{\left(\beta+\frac{\mu}{2}\right)^2-E^{(s)}}$. The purely imaginary poles of $\sigma_0^r(E)$ and $\sigma_0^l(E)$, such that $\kappa^{(0)}>0$ and $\mathcal{\pi}^{(0)}>0$, provide the bosonic bound state eigenvalues and eigenfunctions. Therefore, we need to identify the zeroes of the determinants $\Delta_{0}^r(E,\alpha,\mu,\beta,a)=0=\Delta_{0}^l(E,\alpha,\mu,\beta,a)$ as functions of $E$.
A routine calculation shows that these zeroes are given by the spectral equation:
\begin{equation}
\left(2q\cos 2a q+\mu\sin 2a q\right)\frac{q(\kappa^{(0)}+\mathcal{\pi}^{(0)}+\alpha+\beta)
}{(\kappa^{(0)}+\alpha)(\mathcal{\pi}^{(0)}+\beta)-q^2} =\left(\mu\cos 2a q-2q\sin 2a q\right)-\mu\frac{(\kappa^{(0)}+\alpha)(\mathcal{\pi}^{(0)}+\beta)+q^2}{(\kappa^{(0)}+\alpha)(\mathcal{\pi}^{(0)}+\beta)-q^2} \label{3dbeig} \quad .
\end{equation}
The fermionic bound state eigenvalues are determined from the same equation (\ref{3dbeig}) replacing the $\delta$ walls by wells, i.e., changing $\alpha$, $\mu$, and $\beta$ by $-\alpha$, $-\mu$, and $-\beta$. The unique ground state of the supersymmetric Hamiltonian $\hat{H}$ is fermionic:
\begin{equation*}
\hat{Q}^\dagger\left( \begin{array}
{c}%
0 \\
\psi_{0}^{(1)}(x)\end{array} \right)=0 \, \, \equiv \, \, \hat{D}\psi_{0}^{(1)}(x)=\left[\frac{d}{dx}+\frac{\alpha}{2}\varepsilon(x+a)+\frac{\mu}{2}\varepsilon(x)+\frac{\beta}{2}\varepsilon(x-a)
-\frac{1}{2}(\alpha-\beta)\right]
\psi_{0}^{(1)}(x)=0.
\end{equation*}
The positive energy bosonic bound state, belonging to the discrete spectrum of $\hat{H}_0$, thus complying with the matching conditions,  are built from the scattering solutions for the imaginary external momenta that are poles of the transmission amplitudes.

\subsection{Alternatively attractive and repulsive triple $\delta$-potential of the same strength}
In the case $\alpha=\beta=-\mu$ we find a situation where the steps in the potentials between distinct zones dissappear. The superpotential becomes:
\begin{equation*}
W=\frac{\alpha}{2}|x|-\frac{\alpha}{2}|x-a|-\frac{\alpha}{2}|x+a|.
\end{equation*}
This choice leads to the spectral problem of the general SUSY triple $\delta$-potential restricted to the straight line $(-\alpha,\alpha,-\alpha)$ in the space of parameters $(\alpha,\mu,\beta)\in\mathbb{R}^3$. In this case the potentials are even and the incoming from the left transmission and reflection amplitudes are equal to the analogous amplitudes for scattering waves incoming from the right:
\begin{eqnarray*}
&&  \sigma_1^r(k)=\sigma_1^l(k)=\sigma_1(k) \, \, \, , \, \, \, \rho_1^r(k)=\rho_1^l(k)=\rho_1(k) \quad ; \quad E^{(1)}=k^2+\frac{\alpha^2}{4} \, \, , \\
&&  \sigma_1(k)=\frac{ik+\frac{1}{2}\alpha}{i k-\frac{1}{2}\alpha}\cdot\sigma_0(k) \, \, \, , \, \, \, \rho_1(k)=\frac{-ik-\frac{1}{2}\alpha}{i k-\frac{1}{2}\alpha}\cdot\rho_0(k) \quad ; \quad E^{(0)}=k^2+\frac{\alpha^2}{4} \, \, ,
\end{eqnarray*}
where the identities in the second row between the amplitudes in different sectors come from supersymmetry. From the scattering amplitudes derived above we immediately write down the $\mathbf{S}$-matrices corresponding to $\hat{H}_0$
and $\hat{H}_1$:
\begin{eqnarray*}
&& \hspace{-0.7cm} \mathbf{S}_s =\frac{1}{\Delta_s} 
   \left(\begin{array}{cc}
 8 i k^3 & (-1)^{s+1} 2
   \alpha  [\left(4 k^2+\alpha ^2\right) \cos 2 k a - \alpha ^2-2 k^2 ] \\
(-1)^{s+1}2 \alpha  [\left(4 k^2+\alpha ^2\right) \cos 2 k a - \alpha ^2-2 k^2]
    & 8 i
   k^3
\end{array}
\right)\\ && \Delta_s(k,\alpha, a) =\frac{1}{(2 i
   k+(-1)^s\alpha ) \left[4 k^2+\left(-1+e^{2 i a k}\right)^2 \alpha ^2\right]},
\end{eqnarray*}
where, as usual, $s=0,1$. The poles of the transmission amplitudes $\sigma_0(k)$ and $\sigma_1(k)$ on the positive imaginary half-axis  $k=i\kappa$, $\kappa>0$, in the $k$-complex plane give the bound states respectively of $\hat{H}_0$ and $\hat{H}_1$. The roots of the transcendent equations
\begin{equation}
(2 \kappa-(-1)^s\alpha)\left( 2 \kappa+\alpha -\alpha e^{-2 a \kappa}\right)\left(2\kappa-\alpha+\alpha e^{-2 a \kappa}
   \right)=0;\quad s=0,1, \label{ee3dar0}
\end{equation}
thus characterize the bound state energies and wave functions of both $\hat{H}_0$  and $\hat{H}_1$. There are one or two solutions of (\ref{ee3dar0}) for positive $\kappa$ depending on the relative values of $\alpha$ and $a$. The ground state of the SUSY system is the lower energy bound state of $\hat{H}_0$:
\begin{enumerate}
\item $\kappa_0=\frac{\alpha}{2}$ is a pole of $\sigma_0(k)$ corresponding to the eigenvalue $E_0^{(0)}=0$ of $\hat{H}_0$. Because $\kappa>0$ the bound state wave function $\psi_0(x)$ is normalizable and supersymmetry is unbroken.
\item The second factor in the left member of \eqref{ee3dar0} is only null for $\kappa=0$ giving an unacceptable physical state.
 \item The third factor, however, presents a positive root
$\kappa_1>\kappa_0>0$ of \eqref{ee3dar0} if and only if the distance between $\delta$s is such that $a>\frac{1}{\alpha}$. There is a second bound state of $\hat{H}_0$ of energy $E_1^{(0)}=\frac{\alpha^2}{4}-\kappa_1^2>0=E_0^{(0)}$.
\end{enumerate}
The point spectrum of  $\hat{H}_1$ is also easy to unveil. There is an antibound state of zero energy because the root $\kappa_a=-\frac{\alpha}{2}$ gives a non-normalizable wave function. The other, positive, root of \eqref{ee3dar0} for $s=1$ exists provided that $a>\frac{1}{\alpha}$. This root is exactly the same as the second positive root in \eqref{ee3dar0} for $s=0$. Therefore, there might be only one positive bound state in the point spectrum of $\hat{H}_1$, paired with the second bound state of $\hat{H}_0$, $E_1^{(1)}=E_1^{(0)}>0$, as requested by supersymmetry.
\section{Infinite Dirac $\delta$-interactions of the same strength alternatively attractive and repulsive}\label{s8}
It might be interesting to pursue the analysis of this kind of systems for superpotentials:
\begin{equation}
W(x,\alpha)=\frac{\alpha}{2}\sum_{n=-J}^J\, (-1)^n \vert x- n a \vert \quad , \quad 0<J\in \mathbb{N} \label{supJ}.
\end{equation} 
The essential features of the spectra of these arrays of $N=2J+1$ $\delta$s, however, are captured by the $J=1$ system analysed above. The only possible differences are due to the number of bound states, up to a maximum of $J+1$, depending on the strength $\alpha$ and the distance $a$ between the $\delta$-potentials.

We therefore jump to discuss the $N=+\infty$ array characterized by the superpotential ($\alpha>0$)
\begin{equation*}
W(x)=\frac{\alpha}{2}\sum_{n=-\infty}^\infty(-1)^n\vert x-na \vert \quad , \quad n\in\mathbb{Z}
\end{equation*}
that gives rise to the following Hamiltonians ($s=0,1$):
\begin{equation*}
\hat{H}_s=-\frac{d^{2}}{dx^{2}}+{W^{\prime}}^2+(-1)^s W^{\prime\prime}=-\frac{d^2}{dx^2}+(-1)^s\alpha\sum_{n=-\infty}^\infty(-1)^n\delta(x-na)+\frac{\alpha^{2}}{4}
\end{equation*}
In order to solve the spectral equations for the Schr$\ddot{\rm o}$dinger operators $\hat{H}_0$  and $\hat{H}_1$
\begin{equation*}
-\frac{d^2{\psi}^{(s)}}{dx^2}+V_s(x){\psi}^{(s)}(x)=-\frac{d^2{\psi}^{(0)}}{dx^2}+\left[(-1)^s\alpha\sum_{n=-\infty}^\infty(-1)^n\delta(x-na)
+\frac{\alpha^{2}}{4}\right]\psi^{(s)}(x)=E^{(s)}\psi^{(s)}(x)
\end{equation*}
we emphasize that $V_0$ and $V_1$ are periodic potentials of period $2a$: $V_0(x+2a)=V_0(x)$,
$V_1(x+2a)=V_1(x)$. According to Bloch's theorem the wave functions are quasi-periodic
\be
\psi_q(x+2a)=e^{i 2a q}\psi_q(x) \quad , \quad  -\frac{\pi}{2a}\leq q \leq \frac{\pi}{2a} \, \, ,\nonumber
\ee
the quasi-momentum $q$ characterizing a particular wave function within each (allowed) band.
It is enough to search for the solution in one primitive cell, i.e.,
in the interval $[2na,2(n+1)a]$, where the solution can be written as the linear combination ansatz
$$
\psi_{qn}(x)=\left\{ \begin{array}{ll}
A_n e^{ik(x-2na)}+B_n e^{-ik(x-2na)} & , \ \ 2na<x<(2n+1)a\\
C_n e^{ik(x-2na)}+D_n e^{-ik(x-2na)} & , \ \ (2n+1)a<x<2(n+1)a\\
\end{array} \right.
$$
provided that: $k^2=E-\frac{\alpha^{2}}{4}$. Bloch's theorem now relates the amplitudes between any pair
of primitive cells, e.g., $[0,2a]$ and $[2na,2(n+1)a]$,
in such a way that, $\forall n$ and $k\neq q$,
\be
A_n=e^{2iqna}A_0, \quad \quad \quad B_n=e^{2iqna}B_0 \label{conpc}.
\ee

\subsection{Propagating bands}
In the $[0,2a]$ primitive cell we thus write $\psi_{q0}(x)$ as
\begin{equation}
\psi_{q0}(x)=\left\{ \begin{array}{ll}
A_0 e^{ikx}+B_0 e^{-ikx} & , \ \ 0<x<a\\
C_0 e^{ikx}+D_0 e^{-ikx} & , \ \ a<x<2a
\end{array} \right. \label{wff0} \, ,
\end{equation}
whereas $\psi_{q1}(x)$ in the next $[2a,4a]$ interval reads:
\begin{equation}
\psi_{q1}(x)=e^{2
iqa}\left\{ \begin{array}{ll}
A_0 e^{ik(x-2a)}+B_0 e^{-ik(x-2a)} & , \ \ 2a<x<3a\\
C_0 e^{ik(x-2a)}+D_0 e^{-ik(x-2a)} & , \ \ 3a<x<4a
\end{array} \right. \label{wff1}
\end{equation}
The matching conditions at the point $x=a$, $\psi_{q0}(a_<)=\psi_{q0}(a_>)$ and ${\psi}_{q0}^{\prime}(a_>)-{\psi}_{q0}^{\prime}(a_<)=\pm\alpha\psi_{q0}(a)$ force the two algebraic equations:

\begin{equation}
-e^{i a k}A_0- e^{-i a k}B_0+ e^{i a k}C_0+ e^{-i a k}D_0=0, \label{eq1}
\end{equation}
\begin{equation}
(\mp \alpha  e^{i a k}-i  k e^{i a k})A_0+(i  k e^{-i a k}\mp \alpha  e^{-i a k})B_0+i  k e^{i a k}C_0-i  k e^{-i a k}D_0=0, \label{eq2}
\end{equation}
where the sign $-$ responds to the $V_0$ potential and the sign $+$ arises in the $V_1$ comb.
Two more algebraic equations come from the matching conditions at the boundary point between the two primitive cells: $\psi_{q0}(2a_<)=\psi_{q1}(2a_>)$ and ${\psi}_{q1}^{\prime}(2a_>)-{\psi}_{q0}^{\prime}(2a_<)=\mp\alpha\psi_{q0}(2a)$. With the signs $\pm$ describing respectively the $V_0$ and $V_1$ potentials these equations are:
\begin{equation}
-e^{2 i a q}A_0- e^{2 i a q}B_0+ e^{2 i a k}C_0+ e^{-2 i a k}D_0=0, \label{eq3}
\end{equation}
\begin{equation}
i k e^{2 i a q}A_0-i k e^{2 i a q}B_0+(\pm \alpha  e^{2 i a k}-i  k e^{2 i a k})C_0+(\pm \alpha e^{-2 i a
   k}+i k e^{-2 i a k})D_0=0 \label{eq4} \, .
\end{equation}
All together the four equations \eqref{eq1}, \eqref{eq2}, \eqref{eq3} and
\eqref{eq4} form an homogeneous system of four algebraic equations. The four unknowns $A_0$, $B_0$, $C_0$, and $D_0$ give rise to non-trivial solutions of (\ref{wff0})-(\ref{wff1}) if the determinant of the $4\times 4$-matrix of this algebraic system is zero, i.e., if the spectral equation
\begin{equation}
\cos 2 q a=\frac{(4 k^2+\alpha ^2) \cos 2 k a -\alpha ^2}{4 k^2}, \label{rdkp}
\end{equation}
is satisfied. 

{\bf REMARK}: The determinant of the system and consequently the equation \eqref{rdkp} is invariant under the exchanges of $\alpha\to -\alpha$ and $k\to -k$.
The intersection points of the curves $f(q)=\cos 2 q a$ and $g(k)=\frac{(4 k^2+\alpha ^2) \cos 2 k a -\alpha ^2}{4 k^2}$ determine the correlative values of the quasi-momenta $q$ and the wave momenta $k$
from which solutions of the Schr\"odinger equations for both $\hat{H}_0$ and $\hat{H}_1$ can be built by extending the solutions in one primitive cell to any other primitive cell through use of the connection formulas (\ref{conpc}). 

\begin{figure}[htdp]
\centering
\includegraphics[scale=0.615]{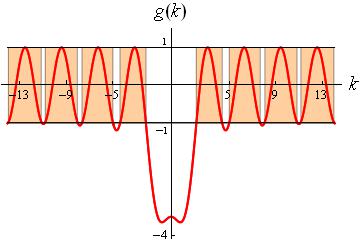}
\caption{ \small Graphics of the curve $g(k)=\frac{1}{4 k^2}(4 k^2+\alpha ^2) \cos 2 k a -\alpha ^2$ for $a=1$ and $\alpha=3$ between $k=0$ and $k=14$ (red). Zones where $g(k)$ varies between $-1$ and $1$, the
minimum and maximum of $f(q)$ (brown). The lower four energy bands are shown. The lower band edges correspond to
$q=\pm\frac{\pi}{2a}+2\pi n$, the upper band edges arise for the quasimomenta $q=2\pi n$. }
\label{rdidar}
\end{figure}

In Figure \ref{rdidar} we show a graphic of the dispersion relation \eqref{rdkp}. Four energy bands are depicted with band lower edges given by the intersection of the $g(k)$ curve with the straight line $y(q)=-1$ whereas the band upper edges are the intersections of $g(k)$ with $y(q)=1$.
Because the invariance with respect to the sign of $\alpha$ these bands are paired by supersymmetry and belongs to both the spectra of $\hat{H}_0$ and $\hat{H}_1$. We also mention that due to the invariance with respect to the sign of $k$ all these wave functions are double degenerate corresponding to propagating waves moving either from left to right or viceversa.

In the allowed bands the solution of the homogeneous system \eqref{eq1}-\eqref{eq4} provides us with the amplitudes $B_0(k)$, $C_0(k)$ and $D_0(k)$ as functions of $A_0(k)$ for any pair $(q,k)$ complying with
\eqref{rdkp}. Plugging these solutions in \eqref{wff0} we obtain the propagating wave functions of the bosonic ($\psi_{q0}^{(0)}$) and fermionic sectors ($\psi_{q0}^{(1)}$) in the
primitive cell $[0,2a]$
{\small \begin{eqnarray*}
&&\left.\psi_{q0}^{(0)}(x)\right\vert_{[0,a]}=\frac{A_0}{N}\left(N e^{i k x}+\left((2 k-i \alpha )e^{4 i a k} -2 k e^{2 i a (k+q)}+i \alpha e^{2 i a k}\right) e^{-i k x} \right)\\
&&\hspace{-0.3cm}\left.\psi_{q0}^{(0)}(x)\right\vert_{[a,2a]}=\frac{A_0}{N}\left(\left((2 k-i \alpha ) e^{2 i a (k+q)}+i \alpha  e^{2 i a q}-2 k\right)  e^{i k x}+\left(i \alpha  e^{2 i a (2k+q)}-(2 k+i \alpha )e^{2 i a (k+q)} +2 k e^{4 i a k}\right) e^{-i k x}\right)
\end{eqnarray*}}
where $A_0=A_0(q,k)$ is a normalization constant, and $N=i
   \alpha  \left(e^{2 i a k}-1\right)-2 k + 2 k e^{2 i a (k+q)}$. The fermionic wave function is obtained from the bosonic one by changing the sign of $\alpha$

\subsection{Non-propagating band}
There may exist also bands of non propagating states if the dispersion relation \eqref{rdkp} admits
solutions with purely imaginary momenta: $k=i\kappa$, $\kappa\in\mathbb{R}$. In the right member of this transcendent equation trigonometric functions are traded by hyperbolic:
\begin{equation}
\cos 2 q a=\frac{(4 \kappa^2- \alpha ^2) \cosh 2 \kappa a +\alpha ^2}{4 \kappa^2}. \label{rdenkp}
\end{equation}
Because the modulus of $g(\kappa)=\frac{(4 \kappa^2-\alpha ^2) \cosh 2 \kappa a +\alpha ^2}{4 \kappa^2}$ is greater than 1 for a relatively low value of $\kappa$ (depending on the values of $a$ and $\alpha$) there is at most one band of this kind of states. As requested by supersymmetry the energy of any state in this band is such that $0\leq E_q < \frac{\alpha^2}{4}$, this last value being the threshold for propagating bands. The bosonic and fermionic non propagating wave functions in the band with energy less than $\frac{\alpha^2}{4}$ have the following form in the primitive cell $[0,2a]$:
{\small\begin{eqnarray}
&&\left.\psi_{q0}^{(0)}(x)\right\vert_{[0,a]}=\frac{A_0}{\widehat N}\left(\widehat N e^{-\kappa x}+\left((2 \kappa - \alpha )e^{-4 a \kappa }-2 \kappa  e^{2 a (i q-\kappa)}+
   \alpha e^{-2 a \kappa }\right)\ e^{\kappa x} \right)\label{npwf0-1}\\
&& \hspace{-1.2cm}\left.\psi_{q0}^{(0)}(x)\right\vert_{[a,2a]}=\frac{A_0}{\widehat N}\left( \left((2  \kappa - \alpha ) e^{2  a (i q- \kappa )}+ \alpha  e^{2 i a q}-2 \kappa \right) \ e^{-\kappa x}
+\left(\alpha  e^{2  a (iq-2\kappa)}-(2 \kappa+ \alpha )e^{2  a (i q-\kappa)}+2 \kappa e^{-4  a \kappa}\right) \ e^{\kappa x} \right)\label{npwf0-2}
\end{eqnarray}}
where $A_0=A_0(q,\kappa)$ is again a normalization constant and $\widehat N=\alpha  \left(e^{-2 a \kappa }-1\right)-2  \kappa+2  \kappa e^{2  a (q-\kappa )}$. In order to determine which are the relative values of $a$ and $\alpha$ such that a given member of the non propagating band characterized by the quasi-momentum $q$ is a truly solution of the Schr\"odinger equation we rewrite equation \eqref{rdenkp} in the form
\begin{equation}
\frac{4 \kappa^2}{\alpha^2}=\frac{1-\cosh 2\kappa a}{\cos 2 q a -\cosh 2\kappa a} \, . \label{rdenkp1}
\end{equation}
We observe that \eqref{rdenkp1}, like \eqref{rdenkp}, is also invariant under $\alpha\to -\alpha$ and $\kappa\to -\kappa$. This means that there is one bosonic and one fermionic state in the non propagating band paired by supersymmetry.
The wave functions obtained from \eqref{npwf0-1} and \eqref{npwf0-2} and the corresponding fermionic ones by applying Bloch's theorem are, however, non-degenerate. The states of the non-propagating bad are descendants of the bound states corresponding to one of the lattice point interactions (see chapter 8 in Ref. \cite{asmer-solid}). Therefore despite the $\kappa\to -\kappa$ invariance of the spectral equation only solutions with $\kappa >0$ are physically accepted \footnote{The states corresponding to solutions of $\kappa<0$ should be interpreted as descendants of the anti-bound states of the single lattice point spectrum}. 
We list the critical relations between $a$ and $\alpha$ for the existence of this kind of solutions:
\begin{itemize}
\item Energy band lower edge solutions: $q=0 \Rightarrow \cos 2qa=1$. \ In this case \eqref{rdenkp1} reduces
to: $\frac{4\kappa^2}{\alpha^2}=1$.
There are two solutions in $\kappa$, $\kappa=\pm\frac{\alpha}{2}$, for any value of $a$ and $\alpha$. We discuss these two solutions separately:
\begin{enumerate}
\item $\kappa=+\frac{\alpha}{2}$. This is the lowest (zero) energy state (being the lower edge in the non-propagating band). Setting $A_0(0,\frac{\alpha}{2}) =1$ the ground state wave function in the primitive cell reads:
\be
\psi_{00}^{(1)}(x)=\left\{ \begin{array}{ll}
e^{-\frac{\alpha}{2}x} & , \ \ 0<x<a\\
e^{\frac{\alpha}{2}x-a \alpha } & , \ \ a<x<2a
\end{array} \right. \label{gswf1}
\ee
We realize that this wave function extended to the whole real line by applying Bloch's theorem is a bona fide zero mode of the Hamiltonian $\hat{H}_1$ under the assumption $\alpha>0$. Note that $V_1(x)$ describes wells at $x=na$ if $n$ is odd but walls for $n$ even. $\psi_{0n}^{(1)}(x)$ exhibits peaks at $x=na$ for both $n$ odd and even but the peaks are maxima at the wells and minima at the walls as it should be. This state behaves badly, however, with respect to $V_0(x)$ because the maxima are at the walls and the minima at the wells, a physically unacceptable situation that must be rejected due to the nature of this state as the analogous to an antibound state arising in the spectrum of $\hat{H}_0$.
\item $\kappa=-\frac{\alpha}{2}$. This state is also the band lower edge of non-propagating wave functions.
It is exactly the (zero mode) ground state of $\hat{H}_0$. In fact, in the primitive cell this wave function reads:
\begin{equation}
\psi_{00}^{(0)}(x)=\left\{ \begin{array}{ll}
e^{\frac{\alpha}{2}x} & , \ \ 0<x<a\\
e^{-\frac{\alpha}{2}x+a \alpha } & , \ \ a<x<2a
\end{array} \right. \label{gswf0} \, .
\end{equation}
The Bloch extension behaves appropriately on the wells (having maxima) and the walls (having minima) of
$V_0(x)$.

\end{enumerate}
To summarise, in the supersymmetric quantum mechanical system there are always one bosonic and one fermionic non-propagating ground states, a paradoxical situation typical of supersymmetric periodic potentials.

\item At upper edge band values of the quasi-momentum $q=\pm\frac{\pi}{2a} \Rightarrow \cos 2qa=-1$ the
equation \eqref{rdenkp1} reduces to the form:
\be
\frac{4 \kappa^2}{\alpha^2}=\frac{\cosh 2\kappa a -1}{\cosh 2\kappa a + 1} \, . \label{rdenkp2}
\ee
Equation \eqref{rdenkp2} is a transcendent one such that the solutions can be only identified by
graphical methods as the intersections of the parabola $f(\kappa)=\frac{4 \kappa^2}{\alpha^2}$ with the
transcendent curve $h(\kappa)=\frac{\cosh 2\kappa a -1}{\cosh 2\kappa a + 1}$. Both curves pass through the origin, $f(0)=0=h(0)$, which is a critical point of $f$ and $h$ as well: $f^\prime(0)=0=h^\prime(0)$. Besides $\kappa=0$, which is a solution that does not correspond to a physical acceptable state the wave function being trivial, the number of intersections depends on the second derivative of these functions at $\kappa=0$: $f^{\prime\prime}(0)=\frac{8}{\alpha^2}$, $h^{\prime\prime}(0)=2 a^2$. There is a critical width $a_c$, depending on the strength $\alpha$, $a_c=\frac{2}{\alpha}$ for which the curvature of the two
curves at the origin coincide. If $a>a_c$ there are one positive $\kappa_+
>0$ and one negative $\kappa_-<0$ solutions of \eqref{rdenkp2}. In this case the whole non-propagating band belongs to the spectrum of the
supersymmetric Hamiltonian, the positive intersection to ${\hat H}_1$ and the negative intersection to
${\hat H}_0$. When $a<a_c$ there are no intersections between $g(\kappa)$ and the straight line $y=-1$: the band upper edge disappears of the ${\hat H}_1$ and ${\hat H}_0$ spectra.

\item The search for solutions of \eqref{rdenkp1} inside the non-propagating band, $-1<\cos 2 q a <1$, also
requires the identification of a critical width. The second derivative of the curve $h(q,\kappa)$ on the right member of this equation is equal to the curvature of $f(\kappa)$ at the $\kappa=0$ point if and only if:
\be
 4 \sin^2 a_c q = \alpha^2 a_c^2 \nonumber \, \, .
\ee
If $a>a_c$ given a solution of this transcendent equation there are two intersections, one positive and the other negative, of the curves in the two members of (36) which means telling that the corresponding wave function belongs to the spectrum of the supersymmetric pair of Hamiltonians. If $a<a_c$ there are no non-null solutions and the non-propagating state disappears from the SUSY spectrum.
\end{itemize}

\begin{figure}[htdp]
\centering
\includegraphics[scale=0.615]{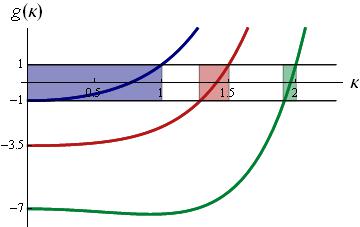}
\caption{ \small Graphics of $g(\kappa)=-\frac{1}{4 \kappa^2}(\alpha ^2-4 \kappa^2) \cosh 2 \kappa a-\alpha ^2$ for $a=1$ and $\alpha=2$ (blue), $\alpha=3$ (brown) and $\alpha=4$ (green). States such that $-1\leq g(\kappa) \leq 1$ form respectively the blue, brown, and green bands.}
\label{rdenidar}
\end{figure}

In Figure \ref{rdenidar} we have depicted the non-propagating band for three different weights $\alpha=2$, $\alpha=3$, $\alpha=4$ and a single width: $a=1$. Note that the last two cases are well within the $a>a_c$ regime
even for the band upper edge. In the first case, however, the width is critical: $a_c=1$. Thus, the band upper edge lies precisely at the boundary.

\subsection{Ground states: the problem of spontaneous supersymmetry breaking}

Analysis of the possible spontaneous symmetry breaking in this system requires us to focus on the characteristics of the ground states. If there are ground states of zero energy such that these states
are singlets of the SUSY algebra supersymmetry is unbroken. The singlet ground states have neccesarily
the form $\psi_0^{(0)}(x)=e^{W(x,\alpha,a)}$ or $\psi_0^{(1)}(x)=e^{-W(x,\alpha,a)}$ respectively in the bosonic and fermionic sectors. In the $J=1$ system $\psi_0^{(0)}(x)=e^{W(x,\alpha,a)}$ is a normalizable state, whereas $\psi^{(1)}(x)\notin L^2(\mathbb{R})$. Thus, there is a bona fide bosonic
singlet ground state and supersymmetry is unbroken. Note that the value of the wave function at the $V_0$ wells, $\psi_0^{(0)}(\pm a)=e^{-\frac{\alpha}{2}a}$, is bigger than the value at the wall: $\psi_0^{(0)}(0)=
e^{-\alpha a}$. It is clear that the wave function exponentially decreases to zero at $\pm\infty$: $\psi_0^{(0)}(\pm\infty)=0$. The reason why we select $\psi_0^{(0)}(x)=e^{W(x,\alpha,a)}$ is also clear. It is not only that changing $W$ by $-W$ makes the wave function non-normalizable but there would be a maximum
at the wall and minima at the wells opposite to expectations from physics. In the next case, $J=2$, we have: $\psi_0^{(1)}(0)=\psi_0^{(1)}(\pm 2a)=e^{-\alpha a}$, $\psi_0^{(1)}(\pm a)=e^{-\frac{3}{2}\alpha a}$, and $\psi_0^{(1)}(\pm \infty)=0$. The situation is identical to the former one in the $J=1$ problem but the normalizable ground state $\psi_0^{(1)}(x)$ belongs to the fermionic sector and $\psi_0^{(0)}(x) \notin L^2(\mathbb{R})$. One can easily check that there is also a single bosonic ground state $\psi_0^{(0)}(x)$ for $J=3$, whereas for $J=4$ the ground state is also unique but fermionic: $\psi^{(1)}(x)$. The general pattern is clear for finite $N=2J+1$: supersymmetry is always unbroken but the \lq\lq regularized\rq\rq Witten index is: $I_W^N=(-1)^{J+1}+1$. It is thus two for $J$ odd and zero for $J$ even. We recall that the addition of $+1$ comes from the difference between the spectral densities in the continuous spectra.

In order to find what happens in the periodic $N\to +\infty$ limit we write in a compact form the superpotential for finite $N=2J+1$ in what is going to be the primitive cell of the periodic potential:
\begin{equation}
W(x,\alpha, a)=\left\{ \begin{array}{ll}
\hspace{0.27cm}\frac{\alpha}{2}(x+2a\sum_{n=1}^J \, (-1)^n n)=\frac{\alpha}{2}\left(x+\frac{a}{2}\left[(-1)^J(1+2J)-1\right]	 \right) & , \ \ 0<x<a \\
-\frac{\alpha}{2}(x-2a\sum_{n=2}^J \, (-1)^n n)=-\frac{\alpha}{2}\left(x-\frac{a}{2}\left[(-1)^J(1+2J)+3\right]\right) & , \ \ a<x<2a \\
\end{array} \right. \label{Nsuppc} \, .
\end{equation}
To take the $J\to \infty$ limit in \eqref{Nsuppc} requires us to consider the Dirichlet eta function defined as the alternating series
\begin{equation*}
\eta(s)=-\sum_{n=1}^\infty\, (-1)^n\cdot\frac{1}{n^s}=-(1-2^{1-s})\zeta(s) \quad , \quad s\in\mathbb{C}
\end{equation*}
at the special point: $s=-1$. Although $\eta(s)$ is strictly convergent only for ${\rm Re}\, s>0$ the relation shown above with the Riemann zeta function $\zeta(s)=\sum_{n=1}^\infty\, \frac{1}{n^s}$ allows us to assign via analytic continuation the value $\eta(-1)=-\frac{1}{4}$ from the sum of all the positive integers: $\zeta(-1)=-\frac{1}{12}$. Therefore, we obtain:
\begin{equation*}
W(x,\alpha, a)=\left\{ \begin{array}{ll}
\frac{\alpha}{2}(x-\frac{1}{2}a) & , \ \ 0<x<a \\
-\frac{\alpha}{2}(x-\frac{3}{2}a) & , \ \ a<x<2a \\
\end{array} \right. \label{Isuppc} \, .
\end{equation*}
Neither $\psi_0^{(0)}(x)=e^{W(x,\alpha, a)}$ nor $\psi_0^{(1)}(x)=e^{-W(x,\alpha, a)}$ are normalizable.
Nevertheless, both wave functions are bona fide ground states. In periodic potentials the wave functions are not normalizable and, moreover, the behaviour of $\psi_0^{(0)}$ and $\psi_0^{(1)}$ properly describes the physics of the problem:
\begin{eqnarray*}
\psi_0^{(0)}(\pm 2n a)= e^{-\frac{\alpha}{4}a} \, \, \, &,& \qquad \psi_0^{(1)}(\pm 2n a)= e^{\frac{\alpha}{4}a}
\\ \psi_0^{(0)}(\pm (2n+1) a)= e^{\frac{\alpha}{4}a} \, \, \, &,& \qquad \psi_0^{(1)}(\pm (2n+1) a)= e^{-\frac{\alpha}{4}a} \, \, .
\end{eqnarray*}
In the $N=\infty$ limit there are two ground states: one bosonic and one fermionic. The Bloch condition
$\psi_q(x+2a)=e^{2iq a}\psi_q(x)$ is satisfied by both
\begin{equation*}
\psi_0^{(0)}(x)={\rm exp}\left[\frac{\alpha}{2}\sum_{n=-\infty}^\infty\, (-1)^n \, \vert x-n a\vert\right] \, \, {\rm and} \, \,
\psi_0^{(1)}(x)={\rm exp}\left[-\frac{\alpha}{2}\sum_{n=-\infty}^\infty\, (-1)^n \, \vert x-n a\vert\right]
\end{equation*}
because $\psi_0^{(0)}(2a)=\psi_0^{(0)}(0)$ and $\psi_0^{(1)}(2a)=\psi_0^{(1)}(0)$. Note that
\begin{eqnarray*}
\psi_0^{(0)}(2a)&=&{\rm exp}\left[\frac{\alpha}{2}a\sum_{n=-\infty}^\infty\, (-1)^n \, \vert 2-n \vert\right] \quad , \quad
\psi_0^{(0)}(0)={\rm exp}\left[\frac{\alpha}{2}a\sum_{n=-\infty}^\infty\, (-1)^n \, \vert -n \vert\right] \\ \psi_0^{(1)}(2a)&=&{\rm exp}\left[-\frac{\alpha}{2}a\sum_{n=-\infty}^\infty\, (-1)^n \, \vert 2-n \vert\right] \quad , \quad
\psi_0^{(1)}(0)={\rm exp}\left[-\frac{\alpha}{2}a\sum_{n=-\infty}^\infty\, (-1)^n \, \vert -n \vert\right]\, \, ,
\end{eqnarray*}
whereas  the two series in the exponents are identical:
\begin{eqnarray*}
\sum_{n=-\infty}^\infty\, (-1)^n \, \vert 2-n \vert&=&2\sum_{j=1}^\infty \, (-1)^j j =2 \eta(-1)=-\frac{1}{2} \, \, , \, \, j=\vert 2-n \vert \\ \sum_{n=-\infty}^\infty\, (-1)^n \, \vert -n \vert&=&2\sum_{n=1}^\infty \, (-1)^n n =2 \eta(-1)=-\frac{1}{2} \, \, \, .
\end{eqnarray*}
Here we have dealt with the Dirichlet eta function $\eta(s)=\sum_{n=1}^\infty (-1)^n\frac{1}{n^s}$ which is finite at $s=-1$. Supersymmetry is thus unbroken but the contribution to the Witten index of the zero energy modes is zero.

\section{Summary and outlook}\label{s9}

We have analyzed in depth the quantum dynamics of one massive particle moving on the real line under the influence of a potential produced by
several configurations of Dirac $\delta$-wells and/or $\delta$-walls. Particular attention has been paid to the study of bound state distribution
as function of the $\delta$-strengths and the distances between them. This information has been used to construct supersymmetric extensions of this system that requires the addition of quasi-square wells. In the supersymmetric framework the dynamics is governed by two intertwined scalar Schr$\ddot{\rm o}$dinger operators with semi-definite spectra: the positive energy eigenfunctions form doublets of the SUSY algebra whereas the zero modes are singlets. We have found that in the supersymmetric systems with a finite number of $\delta$s there is always one zero energy ground state in the spectrum of one of the two partner Hamiltonians such that supersymmetry is unbroken. Letting the number of $\delta$-potentials 
go to infinity spectra of SUSY paired conducting bands arise in the two intertwined operators. The two partner Hamiltonians, however, also admit 
a non-conducting band that encompasses one ground state of zero energy. Supersymmetry is also unbroken but the Witten index is zero.

The concepts, ideas and techniques developed in this paper are prepared to be applied on systems of $N$ particles restricted to move on a line with contact interactions. In particular, the integrable models of Yang, Lieb, and Liniger, see \cite{Yang}-\cite{Lieb}, present a challenge 
of building over them supersymmetric extensions that preserve their integrability. An intriguing question that still to be answered is the possible existence
of more than one ground state of zero energy. One may state, rather vaguely, that in Supersymmetric Quantum Mechanics of one degree of freedom systems only periodic potentials leave room for more than one ground state of zero energy. The $N$-body Yang-Lieb-Liniger systems have $N$ degrees of freedom and one might think in more than one zero energy ground state in the  SUSY version of these systems. In fact, there are two zero energy ground states in the supersymmetric extension of the Euler-Coulomb problem, a charged particle moving on a plane under the action of two Coulombian centres of force, build in Reference \cite{Gonzalez}.  

\section*{Acknowledgement}
J. Mateos Guilarte and J. M. Mu$\tilde{\rm n}$oz Casta$\tilde{\rm n}$eda would like to acknowledge the  support received from the DFG project BO1112-18/1, and the ESF-CASIMIR network.


\section*{References}
\begin{thebibliography}{10}

\bibitem{Albeverio}
P.~Exner and S.~Albeverio.
\newblock {\em Solvable Models in Quantum Mechanics}.
\newblock AMS Chelsea Publishing Series. AMS Chelsea Pub., 2005.

\bibitem{Bordag-jpa25}
Michael Bordag, D.~Hennig, and D.~Robaschik.
\newblock {Vacuum energy in quantum field theory with external potentials
  concentrated on planes}.
\newblock {\em J.Phys.}, A25:4483--4498, 1992.

\bibitem{Lieb}
Elliott~H. Lieb and Werner Liniger.
\newblock {Exact analysis of an interacting Bose gas. 1. The General solution
  and the ground state}.
\newblock {\em Phys.Rev.}, 130:1605--1616, 1963.

\bibitem{Yang}
Chen-Ning Yang.
\newblock {Some exact results for the many body problems in one dimension with
  repulsive delta function interaction}.
\newblock {\em Phys.Rev.Lett.}, 19:1312--1314, 1967.

\bibitem{Cervero}
J.M. Cerver\'o and A.~Rodriguez.
\newblock Infinite chain of different deltas: A simple model for a quantum
  wire.
\newblock {\em The European Physical Journal B - Condensed Matter and Complex
  Systems}, 30(2):239--251, 2002.

\bibitem{Kronig}
R.~de~L. Kronig and W.~G. Penney.
\newblock {Quantum mechanics of electrons in crystal lattices.}
\newblock {\em Proceedings Royal Soc. London (A)}, 130:499--513, 1931.

\bibitem{Brown}
J.S. Guill{\'e}n and M.A. Braun.
\newblock {\em F{\'\i}sica cu{\'a}ntica}.
\newblock Alianza Universidad Textos. Alianza, 1993.

\bibitem{asoreymunoz-npb874}
M.~Asorey and J.M. Munoz-Castaneda.
\newblock {Attractive and Repulsive Casimir Vacuum Energy with General Boundary
  Conditions}.
\newblock {\em Nucl.Phys.}, B874:852--876, 2013.

\bibitem{asoreymunoz-jpa41}
M.~Asorey and J.M. Munoz-Castaneda.
\newblock {Vacuum Boundary Effects}.
\newblock {\em J.Phys.}, A41:304004, 2008.

\bibitem{asoreymunoz-jpa40}
M.~Asorey, D.~Garcia-Alvarez, and J.M. Munoz-Castaneda.
\newblock {Vacuum Energy and Renormalization on the Edge}.
\newblock {\em J.Phys.}, A40:6767--6776, 2007.

\bibitem{Munoz-Castaneda-prd2013}
Jose~M. Munoz-Castaneda, J.~Mateos Guilarte, and A.~Moreno Mosquera.
\newblock {Quantum vacuum energies and Casimir forces between partially
  transparent $\delta$-function plates}.
\newblock {\em Phys.Rev.}, D87:105020, 2013.

\bibitem{aim}
M.~Asorey, A.~Ibort, and G.~Marmo.
\newblock {Global theory of quantum boundary conditions and topology change}.
\newblock {\em Int.J.Mod.Phys.}, A20:1001--1026, 2005.

\bibitem{Witten}
Edward Witten.
\newblock {Dynamical Breaking of Supersymmetry}.
\newblock {\em Nucl.Phys.}, B188:513, 1981.

\bibitem{Wipf}
Andreas Wipf.
\newblock {Non-perturbative methods in supersymmetric theories}.
\newblock {\em Troisieme cycle de la Physique en Suisse Romande}, 2005.

\bibitem{Wipf1}
A.~Kirchberg, J.D. Lange, P.A.G. Pisani, and A.~Wipf.
\newblock {Algebraic solution of the supersymmetric hydrogen atom in
  d-dimensions}.
\newblock {\em Annals Phys.}, 303:359--388, 2003.

\bibitem{Lboya}
L.~J. {Boya}.
\newblock {Quantum mechanical scattering in one dimension}.
\newblock {\em Nuovo Cimento Rivista Serie}, 31(2):020000--139, February 2008.

\bibitem{Boya1}
L~J Boya.
\newblock Supersymmetric quantum mechanics: two simple examples.
\newblock {\em Eur. J. Phys}, 9(2):139, 1988.

\bibitem{Boya2}
L~J Boya, H~C Rosu, A~J Segu\'{\i}-Santonja, J~Socorro, and F~J Vila.
\newblock Supersymmetric one-parameter strict isospectrality for the attractive
  $\delta$ potentials.
\newblock {\em J. Phys. A: Math. Gen.}, 31(44):8835, 1998.

\bibitem{Diaz}
L.~M.~Nieto J.I.~D\'{\i}az, J.~Negro and O.~Rosas-Ortiz.
\newblock The supersymmetric modified p$\ddot{o}$schl-teller and $\delta$-well
  potentials.
\newblock {\em J. Phys. A: Math. Gen.}, 32(48):8447, 1992.

\bibitem{Silva}
F.~Marques, O.~Negrini, and A.J. da~Silva.
\newblock {A new simple class of superpotentials in SUSY Quantum Mechanics}.
\newblock {\em J.Phys.}, A45:115307, 2012.

\bibitem{Plyushchay}
Francisco Correa, Luis-Miguel Nieto, and Mikhail~S. Plyushchay.
\newblock {Hidden nonlinear $su(2|2)$ superunitary symmetry of $N = 2$
  superextended 1D Dirac delta potential problem}.
\newblock {\em Phys.Lett.}, B659:746--753, 2008.

\bibitem{Jakubsky}
Vit Jakubsky, Luis-Miguel Nieto, and Mikhail~S. Plyushchay.
\newblock {The origin of the hidden supersymmetry}.
\newblock {\em Phys.Lett.}, B692:51--56, 2010.

\bibitem{Oeftiger}
A.~Oeftiger.
\newblock {Supersymmetric quantum mechanics}.
\newblock {\em Bachelor Thesis, University of Bern, Switzerland}, pages 1--38,
  2010.

\bibitem{Sukhatme}
Avinash Khare and Uday Sukhatme.
\newblock Periodic potentials and supersymmetry.
\newblock {\em Journal of Physics A: Mathematical and General}, 37(43):10037,
  2004.

\bibitem{Khare2}
Avinash Khare and Uday Sukhatme.
\newblock New solvable and quasiexactly solvable periodic potentials.
\newblock {\em J. Math. Phys.}, 40(11):5473--5494, 1999.

\bibitem{Dunne}
Gerald~V. Dunne and Joshua Feinberg.
\newblock {Self isospectral periodic potentials and supersymmetric quantum
  mechanics}.
\newblock {\em Phys.Rev.}, D57:1271--1276, 1998.

\bibitem{Mannix}
Gerald~V. Dunne and Jake Mannix.
\newblock {Supersymmetry breaking with periodic potentials}.
\newblock {\em Phys.Lett.}, B428:115--119, 1998.

\bibitem{Nieto}
David J.~Fern\'andez C, Javier Negro, and Luis~M. Nieto.
\newblock Second-order supersymmetric periodic potentials.
\newblock {\em Physics Letters A}, 275(5–6):338 -- 349, 2000.

\bibitem{Ioffe}
M.V. Ioffe, J.~Mateos~Guilarte, and P.A. Valinevich.
\newblock {A Class of Partially Solvable Two-Dimensional Quantum Models with
  Periodic Potentials}.
\newblock {\em Nucl.Phys.}, B790:414--431, 2008.

\bibitem{Khare}
Fred Cooper, Avinash Khare, and Uday Sukhatme.
\newblock {Supersymmetry and quantum mechanics}.
\newblock {\em Phys.Rept.}, 251:267--385, 1995.

\bibitem{coop-ap1983}
Fred Cooper and Barry Freedman.
\newblock {Aspects of Supersymmetric Quantum Mechanics}.
\newblock {\em Annals Phys.}, 146:262, 1983.

\bibitem{cromb-ap1983}
M.~de~Crombrugghe and V.~Rittenberg.
\newblock {Supersymmetric Quantum Mechanics}.
\newblock {\em Annals Phys.}, 151:99, 1983.

\bibitem{cooper-pla1988}
Fred Cooper, Joseph~N. Ginocchio, and Andreas Wipf.
\newblock {Derivation of the S Matrix Using Supersymmetry}.
\newblock {\em Phys.Lett.}, A129:145--147, 1988.

\bibitem{Witten1}
Edward Witten.
\newblock {Constraints on Supersymmetry Breaking}.
\newblock {\em Nucl.Phys.}, B202:253, 1982.

\bibitem{Freund}
P.G.O. Freund.
\newblock {\em Introduction to Supersymmetry}.
\newblock Cambridge Monographs on Mathematical Physics. Cambridge University
  Press, 1988.

\bibitem{for-npb1987}
P.~Forgacs, L.~O'Raifeartaigh, and A.~Wipf.
\newblock {Scattering Theory, U(1) Anomaly and Index Theorems for Compact and
  Noncompact Manifolds}.
\newblock {\em Nucl.Phys.}, B293:559, 1987.

\bibitem{fucci-spp2011}
Guglielmo Fucci, Klaus Kirsten, and Pedro Morales.
\newblock {Pistons Modelled by Potentials}.
\newblock {\em Springer Proc.Phys.}, 137:313--322, 2011.

\bibitem{fucci-ijmp2012}
Guglielmo Fucci and Klaus Kirsten.
\newblock {The Casimir Effect for Generalized Piston Geometries}.
\newblock {\em Int.J.Mod.Phys.Conf.Ser.}, 14:100--114, 2012.

\bibitem{kirsten-prd09}
Klaus Kirsten and S.A. Fulling.
\newblock {Kaluza-Klein models as pistons}.
\newblock {\em Phys.Rev.}, D79:065019, 2009.

\bibitem{asmer-solid}
N.W. Ashcroft and N.D. Mermin.
\newblock {\em Solid State Physics}.
\newblock Cengage Learning, 2011.

\bibitem{Gonzalez}
M.A. Gonzalez~Leon, M.~de~la Torre~Mayado, J.~Mateos~Guilarte, and M.J.
  Senosiain.
\newblock {On the Supersymmetric Spectra of two Planar Integrable Quantum
  Systems}.
\newblock {\em Contemporary Mathematics}, 563:73, 2012.

\end{thebibliography}
\end{document}